\documentclass[nofootinbib,reprint,amsmath,amssymb,aps]{revtex4-1}
\usepackage{graphicx}
\usepackage{dcolumn}
\usepackage{bm}

\usepackage{hyperref}
\hypersetup{colorlinks, linkcolor={red},citecolor={blue},urlcolor={blue}}  
\newcommand\ChangeRT[1]{\noalign{\hrule height #1}}

\begin{document}

\preprint{APS/123-QED}

\title[Three-dimensional general relativistic Poynting-Robertson effect. III]{Three-dimensional general relativistic Poynting-Robertson effect. III.\\ Static and non-spherical quadrupolar massive source}

\author{Vittorio De Falco$^{1}$}\email{vittorio.defalco@physics.cz}
\author{Pavel Bakala$^{1,2}$}\email{pavel.bakala@fpf.slu.cz}
\author{Maurizio Falanga$^{3}$} 
\vspace{0.5cm}

\affiliation{$^1$M. R. \v{S}tef\'anik Observatory and Planetarium, Sl\'adkovi\v{c}ova 41, 920 01 Hlohovec, Slovak Republic\\
$^2$ Research Centre for Computational Physics and Data Processing, Faculty of Philosophy \& Science, Silesian University in Opava, Bezru\v{c}ovo n\'am.~13, CZ-746\,01 Opava, Czech Republic\\
$^3$ International Space Science Institute, Hallerstrasse 6, 3012 Bern, Switzerland}

\date{\today}

\begin{abstract}
We investigate the three-dimensional (3D) motion of a test particle in the gravitational field generated by a non-spherical compact object endowed with a mass quadrupole moment, described by the Erez-Rosen metric, and a radiation field, including the general relativistic Poynting-Robertson (PR) effect, coming from a rigidly rotating spherical emitting source located outside of the compact object. We derive the equations of motion for test particles influenced by such radiation field, recovering the two-dimensional (2D) description, and the weak-field approximation. This dynamical system admits the existence of a critical hypersurface, region where gravitational and radiation forces balance. Selected test particle orbits for different set of input parameters are displayed. The possible configurations on the critical hypersurfaces can be either  latitudinal drift towards the equatorial ring or suspended orbits. We discuss about the existence of multiple hypersurface solutions through a simple method to perform the calculations. We graphically prove also that the critical hypersurfaces are stable configurations within the Lyapunov theory. 
\end{abstract}

\maketitle
\section{Introduction}
\label{sec:intro}
In high-energy astrophysics the motion of matter around a compact object, as a black hole (BH) or a neutron star (NS), can be influenced not only by the gravitational field, but also by electromagnetic radiation forces. The radiation pressure is directed radially outward from an emitting source to the matter position, contrasting thus the enormous gravitational pull from the compact object. Beside that, there is also the PR effect, pure relativistic dissipative force, which efficiently removes energy and angular momentum from the affected body \cite{Poynting1903,Robertson1937}.  

In 2009 -- 2011, Bini and collaborators presented for the first time its general relativistic treatment in the 2D equatorial plane in Kerr metric \cite{Bini2009,Bini2011}, and only recently it has been extended in the 3D case \cite{DeFalco20183D,Bakala2019,Wielgus2019}. Such effect has been also treated under a Lagrangian formalism, determining for the first time the analytical form of the Rayleigh potential \cite{DeFalco2018,DeFalco2019L,DeFalco2019VE}. All models of the general relativistic PR effect share a common propriety, namely the existence of a critical hypersurface around the compact object, where gravitational attraction and radiation force balance. It has been also formally proven within the Lyapunov theory that the equatorial ring of the critical hypersurface is a stable attractor, and the whole critical hypersurface is a basin of attraction \cite{DeFalco2019ST}.

Realistic astrophysical bodies possess a certain set of multipole moments. Actually, even extremely compact bodies like NSs or BHs are not spherical in general. Under the assumption of stationarity, they should be axially symmetric, but, due to rotation and/or a possible presence of external matter or fields, they may have non-negligible dipole and quadrupole moments. Restricting further to just a \emph{static} case, thus neglecting dragging effects due to the rotation, one is left with monopole and quadrupole. Such a deformed static body can be described by the Erez-Rosen metric, which is an exact solution of Einstein's field equations in the vacuum endowed with a quadrupole moment $q\in\mathbb{R}$ and a mass $M$ (see Refs. \cite{Armenti1977,Quevedo1989,Quevedo1990,Quevedo1991,Krori1991,Mashhoon1995}, for more details) can better describe non-spherical massive compact objects.

This feature is relevant in high-energy astrophysics, indeed a non-negligible mass quadrupole moment significantly affects the dynamics of close surrounding compact objects for the presence of additional tidal forces. There is a strong evidence in favor of such hypothesis thanks to the advanced technology in astrometry to accurately monitor for many years the compact cluster of stars orbiting the centre of the Milky Way galaxy at milliparsec distances \cite{Merritt2010,Gillessen2017}. To this end, Johannsen \& Psaltis have developed a parametric framework for testing the no-hair theorem in terms of the quadrupole moment against the observations in the electromagnetic spectrum \cite{Johannsen20101,Johannsen20102,Johannsen2011,Johannsen2013}. However along such research line, there is also another more recent parametrization due to Konoplya, Rezzolla, and Zhidenko \cite{Konoplya2016}. 

In this work, we aim to extend the work of Ref. \cite{Bini2015} in the 3D case, using as a description of the radiation field the model developed in Ref. \cite{Bakala2019}. The paper is organized as follows: in Sec. \ref{sec:geometry} we describe the geometrical environment in which our problem is set, motivating also the choice of the Eerez-Rosen metric; in Sec. \ref{sec:dynamics} we recall the radiation field model, and then derive the general relativistic 3D equations of motion; in Sec. \ref{sec:CH} we analyse the critical hypersurfaces, investigating extensively their proprieties, and displaying also some test particle trajectories, and graphically proving their stability in Lyapunov theory; finally in Sec. \ref{sec:end} we draw our conclusions. 

\section{Spacetime geometry}
\label{sec:geometry}
We study the motion of a test particle orbiting around a static and non-spherical compact object in General Relativity (GR). Outside the compact object there is a spherical rigidly rotating emitting surface producing a radiation field, which includes a radiation pressure, opposite to the gravitational field, and the general relativistic PR effect. This section focuses on the geometrical description of the problem, and it is organized as follows: in Sec. \ref{sec:gen_met} we describe the Weyl class of static and axisymmetric solutions of vacuum Einstein field equations to describe non-spherical and static compact objects; in Sec. \ref{sec:ER_met} we describe the features of the Erez-Rosen metric; in Sec. \ref{sec:obsrver} the local static observers are introduced; in Sec. \ref{sec:kin} we define and calculate all the kinematical quantities involved in this framework. 

\subsection{Static and axially symmetric spacetime metrics around non-spherical massive source}
\label{sec:gen_met}
We are interested in describing the spacetime around a static and non-spherical massive source through an exact vacuum solution of the Einstein's field equations. Therefore, a reasonable metric must be expressed in terms of mass multipole moments. The Schwarzschild solution represents the first exact solution of Einstein's field equations in empty space with only monopole, representing the total conserved mass-energy contained in a body \cite{Schwarzschild1916}. In 1917 Weyl and Levi-Civita have found a \emph{class of static and axially symmetric solutions to the vacuum Einstein's field equations} \cite{Weyl1918,Weyl1919,Levi1919}. A suitable way to describe the most general line element for this type of gravitational field is through the cylindrical coordinates $(t,\rho,z,\varphi)$ in the Weyl-Lewis-Papapetrou form \cite{Weyl1918,Lewis1932,Papapetrou1953,Stephani2003,Frutos2018} as
\begin{equation} \label{eq:weyl_metric}
ds^2=-e^{2\mu}dt^2+e^{-2\mu}[e^{2\lambda}(d\rho^2+dz^2)+\rho^2d\varphi^2],
\end{equation}
where $\mu=\mu(\rho,z)$ and $\lambda=\lambda(\rho,z)$, which satisfies, through the vacuum Einstein field equations, the following set of independent partial differential equations:
\begin{equation} \label{eq:weyl_condition}
\begin{aligned}
&\partial_{\rho\rho}\mu+\frac{1}{\rho}\partial_\rho \mu+\partial_{zz}\mu=0,\\
&\partial_\rho\lambda=\rho(\partial_\rho^2-\partial_z^2),\quad \partial_z\lambda=2\partial_\rho\mu \partial_z\mu.
\end{aligned}
\end{equation}
Weyl proposed also a general solution with the further request of asymptotic flatness \cite{Weyl1918,Frutos2018}. A particular solution has to admit as limiting case a pure spherical massive source described by the Schwarzschild's spherically symmetric spacetime. To investigate the proprieties of such solutions with multipole moments, it has been noted that it is more convenient to use \emph{prolate spheroidal coordinates} $(t, x, y, \varphi)$ \cite{Frutos2018}, where the transformation with the previous coordinate system is given by 
\begin{equation} \label{eq:coordinate}
\begin{aligned}
&x=\frac{r_++r_-}{2M}\quad (x^2\ge1),\\
&y=\frac{r_+-r_-}{2M}\quad (y^2\le1),\\
&r_\pm^2=\rho^2+(z\pm M)^2,
\end{aligned}
\end{equation}
where $M$ represents the constant total mass-energy of the body and $\mu=\mu(x,y)$ and $\lambda=\lambda(x,y)$. In such coordinate system, the line element can be written as 
\begin{equation} \label{eq:psc_metric}
\begin{aligned}
ds^2&=-e^{2\mu}dt^2+\frac{M^2}{e^{2\mu}}\left\{e^{2\lambda}(x^2-y^2)\left(\frac{dx^2}{x^2-1}+\frac{dy^2}{1-y^2}\right)\right.\\
&\left.+(x^2-1)(1-y^2)d\varphi^2\right\},
\end{aligned}
\end{equation}
and the Einstein field equations become
\begin{equation} \label{eq:psc_condition}
\begin{aligned}
&\partial_x[(x^2-1)\partial_x\mu]+\partial_y[(1-y^2)\partial_y\mu]=0,\\
&\partial_x\lambda=\left(\frac{1-y^2}{x^2-y^2}\right)\left[x(x^2-1)\partial_x\mu^2\right.\\
&\left.\qquad-x(1-y^2)\partial_y\mu^2-2y(x^2-1)\partial_x\mu\partial_y\mu\right],\\
&\partial_y\lambda=\left(\frac{x^2-1}{x^2-y^2}\right)\left[y(x^2-1)\partial_x\mu^2\right.\\
&\left.\qquad-y(1-y^2)\partial_y\mu^2-2x(1-y^2)\partial_x\mu\partial_y\mu\right],\\
\end{aligned}
\end{equation}
In order to find a \emph{particular physically meaningful solution}, it is necessary that the metric satisfies the following conditions: ($i$) \emph{asymptotic flatness} (it reduces to the Minkowski metric at spatial infinity), ($ii$) \emph{elementary flatness} (it admits no conical singularities on the axis) and ($iii$) \emph{regularity} (it must be free of curvature singularities outside a region located near the origin of coordinates, so that it can be covered by an interior solution) \cite{Frutos2018}. To determine a specific metric, it must be given the explicit expression of $\mu$ e $\lambda$ as solutions of Eqs (\ref{eq:psc_condition}). Over the course of time,  different exact solutions of the Weyl equation have been proposed in the literature, i.e.: Erez \& Rosen (1959) \cite{Erez1959}, Gutsunayev \& Manko (1985) \cite{Gutsunayev1985}, Manko (1990) \cite{Manko1990}, Hern\'andez-Pastora \& Mart\'in (1994) \cite{Hernandez1994}. In 1966 -- 1970 Zipoy and Voorhees found a transformation, based on a particular symmetry of the Weyl equations, which can be used to generate new solutions from known solutions \cite{Zipoy1966,Voorhees1970}. All the solutions are expressed in terms of the mass $M$ and of the additional dimensionless quadrupole parameter $q$. All of them reduce to the Schwarzschild metric in the limiting case $q\rightarrow0$ through the following transformation 
\begin{equation}
x=\frac{r}{M}-1,\qquad y=\cos\theta, 
\end{equation}
where $r$ and $\theta$ are respectively the radius and latitudinal angle of spherical coordinates. From a physical point of view, the most important multipoles of a non-rotating (but also for a rotating) mass distribution are represented by monopole (or simple total mass of the system) and quadrupole moments. Therefore, it arises spontaneous to question: \emph{in the Weyl class, which is the appropriate metric to describe the geometry around a non-spherical massive source endowed only with monopole and quadrupole moments?} It has been proved that all of them are equivalent up to the quadrupole moment approximation through a simple redefinition of the parameter $q$. \emph{Therefore, all the above mentioned metrics exhibit equivalent values of the monopole and quadrupole moments, and valuable differences can appear only at higher multipole moments} (see Ref. \cite{Frutos2018}, for further details). 

\subsection{Erez-Rosen spacetime}
\label{sec:ER_met}
Due to this equivalence result up to the quadrupole moment, we decide to consider the Erez \& Rosen metric \cite{Erez1959}, taking advantage of some calculations already developed in Ref. \cite{Bini2015}. Such metric was corrected for several numerical coefficients by Doroshkevich and collaborators \cite{Doroshkevich1966}, and Young and Coulter \cite{Young1969}. The line element $ds^2=g_{\alpha\alpha}(dx^\alpha)^2$, written in prolate spheroidal coordinates $(t, x, y, \varphi)$, is given by Eq. (\ref{eq:psc_metric}), and
\begin{equation} \label{eq:coeff}
\begin{aligned}
\mu&=-Q_0-q P_2 Q_2,\\
\lambda&=\frac{1}{2}(1+q)^2\ln\left(\frac{x^2-1}{x^2-y^2}\right)+q \Psi,\\
\Psi&=2(1-P_2)Q_1+q(1-P_2)\left[(1+P_2)(Q_1^2-Q_2^2)\right.\\
&\left.+\frac{1}{2}(x^2-1)\mathcal{A}\right],\\
&\mathcal{A}=2Q_2^2-3xQ_1Q_2+3Q_0Q_2-\partial_xQ_2,
\end{aligned}
\end{equation}
where $P_l(y)$ and $Q_l(x)$ are the $l$th Legendre polynomial of the first and second kind\footnote{The Legendre polynomials used in the metric are:
\begin{equation}\label{eq:LP}
\begin{aligned}
&Q_0(x)=\frac{1}{2}\ln\left(\frac{x+1}{x-1}\right),\ Q_1(x)=xQ_0(x)-1,\\
&Q_2(x)=-\frac{1}{2}[Q_0(x)-3xQ_1(x)],\ P_2(y)=-\frac{1}{2}(1-3y^2).
\end{aligned}
\end{equation}
}, respectively \cite{LPl964}. The determinant of the metric is $\sqrt{-g}=M^3e^{2(\lambda-\mu)}(x^2-y^2)$. This spacetime departs significantly from the Schwarzschild metric for the presence of the quadrupole parameter $q$. For $q>0$ the mass is concentred along the $y$-axis (prolate configurations), instead for $q<0$ the mass is distributed along the $x$-axis (oblate configurations) \cite{Bini2015}. Due to the symmetry of the background spacetime, we have that $\boldsymbol{\partial_t}$ (timelike) and $\boldsymbol{\partial_\varphi}$ (spacelike) are commuting Killing vectors. Since the compact object is not rotating, the shift vector field $N^\varphi=0$, and there is only the lapse function $N\equiv(-g_{tt})^{-1/2}=e^{\mu}$ \cite{Jantzen1992,Bini1997a,Bini1997b,DeFalco2018}.

\subsection{Local static observers}
\label{sec:obsrver}
Since there is no frame dragging effects on the spacetime background, a suitable family of fiducial observers is represented by local static observers, with unit timelike four-velocity $\boldsymbol{n}=N^{-1}\boldsymbol{\partial_t}$ aligned with the timelike Killing vector $\boldsymbol{\partial_t}$. These observers to be static must move against the gravitational pull, therefore they are endowed with acceleration $\boldsymbol{a}(\boldsymbol{n})=\nabla_{\boldsymbol{n}}\boldsymbol{n}$ \cite{Misner1973}. An orthonormal frame adapted to the local static observers is given by
\begin{equation}\label{eq:ad_frame}
\begin{aligned}
&\boldsymbol{e_{\hat t}}=\boldsymbol{n},\
\boldsymbol{e_{\hat x}}=\frac{\boldsymbol{\partial_x}}{\sqrt{g_{xx}}},\
\boldsymbol{e_{\hat y}}=\frac{\boldsymbol{\partial_y}}{\sqrt{g_{yy}}},\
\boldsymbol{e_{\hat \varphi}}=\frac{\boldsymbol{\partial_\varphi}}{\sqrt{g_{\varphi\varphi}}}.
\end{aligned}
\end{equation}
All the indices associated with tensorial or vectorial quantities (i.e., $v^\alpha,T^{\alpha\beta}$) in the local static observer frame will be labeled by a hat index (i.e., $v^{\hat\alpha},T^{\hat{\alpha}\hat{\beta}}$), instead all the scalar quantities (i.e., $f$) measured in the local static observer frame will be followed by $(\boldsymbol{n})$ (i.e., $f(\boldsymbol{n})$). 

\subsection{Kinematical quantities}
\label{sec:kin}
The local static observers are globally non rotating and not expanding, which respectively implies that their vorticity tensor $\boldsymbol{\omega}(\boldsymbol{n})$ and their expansion tensor $\boldsymbol{\theta}(\boldsymbol{n})$ vanish. In this case, the Lie transport coincides with the Fermi-Walker transport (see \cite{Bini1997a,Bini1997b,DeFalco2018}, for further details). The non-negative kinematical quantities are the four acceleration $\boldsymbol{a}(\boldsymbol{n})$ and the signed Lie curvature tensors $\boldsymbol{k}(x^i,\boldsymbol{n})$ relative to the local static observer timelike velocity $\boldsymbol{n}$, representing the coordinate line curvatures of the test particle trajectory respectively in the directions of $x^i=x,y,\varphi$ (see \cite{Bini1997a,Bini1997b,DeFalco2018}, for further details). They all have non-zero components in the $\boldsymbol{\hat{x}}-\boldsymbol{\hat{y}}$ tangent plane to the local static observers \cite{Bini2015}, i.e.,
\begin{equation}\label{eq:kinematics}
\begin{aligned}
\boldsymbol{a}(\boldsymbol{n})&=a(\boldsymbol{n})^{\hat x}\ \boldsymbol{e_{\hat x}}+a(\boldsymbol{n})^{\hat y}\ \boldsymbol{e_{\hat y}}\\
&=\frac{\partial_x(\ln N)}{\sqrt{g_{xx}}}\ \boldsymbol{\partial_{x}}+\frac{\partial_y(\ln N)}{\sqrt{g_{yy}}}\ \boldsymbol{\partial_{y}},\\
\boldsymbol{k}(x^i,\boldsymbol{n})&=k(x^i,\boldsymbol{n})^{\hat x}\ \boldsymbol{e_{\hat x}}+k(x^i,\boldsymbol{n})^{\hat y}\ \boldsymbol{e_{\hat y}}\\
&=-\frac{\partial_x(\ln\sqrt{g_{ii}})}{\sqrt{g_{xx}}}\ \boldsymbol{\partial_{x}}-\frac{\partial_y(\ln\sqrt{g_{ii}})}{\sqrt{g_{yy}}}\ \boldsymbol{\partial_{y}},
\end{aligned}
\end{equation}
We extend our analysis in the 3D spacetime, namely outside of the symmetric equatorial plane $y=0$. All the quantities directed along the $x$-axis will be termed from now on as \emph{radial}, while those directed along the $y$-axis will be termed as \emph{polar}. In Table \ref{tab:SOq} the explicit expressions of such quantities are calculated.
\renewcommand{\arraystretch}{1.8}
\begin{table*}[t!]
\begin{center}
\caption{\label{tab:SOq} Explicit expressions of metric and local static observer kinematical quantities for the Erez-Rosen metric.\\}	
\normalsize
\begin{tabular}{l  c} 
\ChangeRT{1pt}
{\bf Metric quantity} & {\bf Explicit expression} \\
\ChangeRT{1pt}
$Q_0'\equiv \partial_x Q_0$ & $-\frac{1}{x^2-1}$\\
\hline
$Q_1'\equiv \partial_x Q_1$ & $Q_0+xQ_0'$\\
\hline
$Q_2'\equiv \partial_x Q_2$ & $Q'_0+3Q_1$\\
\hline
$Q_2''\equiv \partial_x Q_2'$ & $2x(Q'_0)^2+3Q_1'$\\
\hline
$\mathcal{A}'\equiv \partial_x \mathcal{A}$ & $4Q_2Q_2'-3Q_1Q_2-3xQ_1'Q_2-3xQ_1Q_2'+3Q_0'Q_2+3Q_0Q_2'-Q_2''$\\
\hline
$\partial_x\Psi$ & $2(1-P_2)Q_1'+q(1-P_2)[2(1+P_2)(Q_1Q_1'-Q_2Q_2')+x\mathcal{A}+\frac{1}{2}(x^2-1)\mathcal{A}']$\\
\hline
$\partial_y\Psi$ & $-6y\left\{Q_1+q\left[P_2(Q_1^2-Q_2^2)+\frac{1}{4}(x^2-1)\mathcal{A}\right]\right\}$\\
\hline
$\partial_x\mu$ & $-Q'_0-qP_2Q_2'$\\
\hline
$\partial_y\mu$ & $-3yqQ_2$\\
\hline
$\partial_x\lambda$ & $(1+q)^2\frac{x(1-y^2)}{(x^2-y^2)(x^2-1)}+q\partial_x\Psi$\\
\hline
$\partial_y\lambda$ & $(1+q)^2\frac{y}{x^2-y^2}+q\partial_y\Psi$\\
\ChangeRT{1pt}
{\bf Kinematical quantity} & {\bf Explicit expression} \\
\ChangeRT{1pt}
\hline
\hline
\multicolumn{2}{c}{\emph{Radial components}}\\
\hline
\hline
$a(\boldsymbol{n})^{\hat x}$&  $\frac{1}{\sqrt{g_{xx}}}\left(\frac{1}{x^2-1}-qP_2Q_2'\right)$\\
\hline
$k(x,\boldsymbol{n})^{\hat x}$ & $-\frac{1}{\sqrt{g_{xx}}}\left[\partial_x\lambda-\partial_x\mu-\frac{x(1-y^2)}{(x^2-1)(x^2-y^2)}\right]$ \\
\hline
$k(y,\boldsymbol{n})^{\hat x}$ & $-\frac{1}{\sqrt{g_{xx}}}\left[\partial_x\lambda-\partial_x\mu+\frac{x}{x^2-y^2}\right]$\\
\hline
$k(\varphi,\boldsymbol{n})^{\hat x}$ &$-\frac{1}{\sqrt{g_{xx}}}\left[-\partial_x\mu+\frac{x}{x^2-1}\right]$ \\
\hline
\hline
\multicolumn{2}{c}{\emph{Polar components}}\\
\hline
\hline
$a(\boldsymbol{n})^{\hat y}$& $-\frac{3qyQ_2}{\sqrt{g_{yy}}}$ \\
\hline
$k(x,\boldsymbol{n})^{\hat y}$ & $-\frac{1}{\sqrt{g_{yy}}}\left[\partial_y\lambda-\partial_y\mu-\frac{y}{x^2-y^2}\right]$\\
\hline
$k(y,\boldsymbol{n})^{\hat y}$ & $-\frac{1}{\sqrt{g_{yy}}}\left[\partial_y\lambda-\partial_y\mu+\frac{y(x^2-1)}{(1-y^2)(x^2-y^2)}\right]$\\
\hline
$k(\varphi,\boldsymbol{n})^{\hat y}$ & $\frac{1}{\sqrt{g_{yy}}}\left[\partial_y\mu+\frac{y}{1-y^2}\right]$\\
\ChangeRT{1pt}
\end{tabular}
\end{center}
\end{table*}

\section{Test particle dynamics in Erez-Rosen spacetime under the 3D general relativistic PR effect}
\label{sec:dynamics}
In this section, we describe how to treat the radiation field and derive the equations of motion. In Sec. \ref{sec:radfield} we introduce the radiation field model; in Sec. \ref{sec:tpmotion} we describe the test particle motion, including velocity and acceleration fields; in Sec. \ref{sec:radfield} we deal with the test particle-radiation field interaction; in Sec. \ref{sec:eqm} the equations of motion are ensued by these premises. 

\subsection{Radiation field}
\label{sec:radfield}
The radiation field is modeled by photons traveling along null geodesics of the Erez-Rosen spacetime with the energy tensor given by \cite{Bini2015,DeFalco2018,DeFalco20183D,Bakala2019}
\begin{equation} \label{eq:radfield}
T^{\alpha\beta}=\Phi^2 k^\alpha k^\beta,\quad k^\alpha k_\alpha=0,\quad k^\beta\nabla_\beta k^\alpha=0.
\end{equation}
The photon four-momentum $\boldsymbol{k}$ can be split in the local static observer frame as \cite{Bini2015,DeFalco2018,DeFalco20183D,Bakala2019}
\begin{eqnarray} 
&&\boldsymbol{k}=E(\boldsymbol{n})[\boldsymbol{n}+\boldsymbol{\hat{\nu}} (\boldsymbol{k},\boldsymbol{n})],\label{eq:ksplit}\\
&&\boldsymbol{\hat{\nu}}(\boldsymbol{k},\boldsymbol{n})=\sin\beta\sin\xi\ \boldsymbol{e_{\hat x}}+\cos\xi\ \boldsymbol{e_{\hat y}}+\sin\xi\cos\beta\ \boldsymbol{e_{\hat \varphi}},\notag
\end{eqnarray}
where $\boldsymbol{\hat{\nu}} (\boldsymbol{k},\boldsymbol{n})$ is the photon spatial velocity on the spatial hypersurface orthogonal to $\boldsymbol{n}$, and $E(\boldsymbol{n})$ is the relative photon energy in the local static observer frame \cite{Bini2015}
\begin{equation} \label{eq:ene}
E(\boldsymbol{n})=-\boldsymbol{k}\cdot\boldsymbol{n}=-\boldsymbol{k}\cdot\frac{\boldsymbol{\partial_t}}{N}=\frac{E}{e^{\mu}},
\end{equation}
where $E=-k_t>0$ is the conserved photon energy, $\beta$ and $\xi$ are the two angles in the azimuthal and polar direction, respectively. The case $\sin\beta > 0$ corresponds to outgoing photons (increasing radial distance from the central source), and $\sin\beta < 0$ to incoming photons (decreasing $x$). The angular momentum along the polar $\boldsymbol{\hat y}$-axis in the local static observer frame, $L_{\hat y}(\boldsymbol{n})$ is \cite{DeFalco20183D,Bakala2019}
\begin{equation} \label{eq:angmom}
\begin{aligned}
E(\boldsymbol{n})\cos\beta\sin\xi&=L_{\hat y}(\boldsymbol{n})=
\boldsymbol{k}(\boldsymbol{n})\cdot \boldsymbol{e_{\hat \varphi}}\\
&=\boldsymbol{k}\cdot \frac{\boldsymbol{\partial_\varphi}}{\sqrt{g_{\varphi\varphi}}}=\frac{L_y}{\sqrt{g_{\varphi\varphi}}},
\end{aligned}
\end{equation}
where $L_y = k_\varphi$ is the conserved photon angular momentum along the $\boldsymbol{y}$-axis. From Eqs. (\ref{eq:ene}) and (\ref{eq:angmom}), we have  
\begin{equation} \label{eq:ang}
\cos\beta=\frac{be^{2\mu}}{\sin\xi M\sqrt{(x^2-1)(1-y^2)}},
\end{equation}
where $b = L_y/E$ denotes the azimuthal photon impact parameter associated to the azimuthal angle $\beta$. An equation for the latitudinal angle $\xi$ is necessary to completely determine $\beta$. The specific photon four-momentum components in the Erez-Rosen geometry are \cite{Quevedo1991,Bini2013}
\begin{eqnarray}\label{eq:nullgeo}
k^t&&=-\frac{1}{e^{2\mu}},\notag\\
k^x&&=\frac{\sqrt{\sin^2\xi M^2(x^2-1)(1-y^2)-b^2 e^{4\mu}}}{M e^{\lambda}}\sqrt{\frac{1-y^2}{x^2-y^2}},\notag\\
k^y&&=\frac{\cos\xi}{e^\mu Me^{(\lambda-\mu)}}\sqrt{\frac{1-y^2}{x^2-y^2}},\\
k^\varphi&&=\frac{b}{M^2 e^{-2\mu}(x^2-1)(1-y^2)}.\notag
\end{eqnarray}
We consider a radiation field emitted from a spherical and rigidly rotating emitting source (see Ref. \cite{Bakala2019}, for further details), where everything is expressed in terms of the only parameter $b$. 
This implies that $\xi=\pi/2$, and for $x_\star=R_\star/M-1$ we have
\begin{equation}
b=\left[-\frac{g_{\varphi\varphi}}{g_{tt}}\right]_{x=x_\star}\Omega_\star,
\end{equation}
where $R_\star$ and $\Omega_\star$ are respectively radius and angular velocity of the emitting surface. Therefore, in view of these results, we have that the azimuthal photon angle in the local static observer frame (\ref{eq:ang}) becomes
\begin{equation} \label{eq:ang2}
\cos\beta=\frac{be^{2\mu}}{M\sqrt{(x^2-1)(1-y^2)}}.
\end{equation}

Since the photon four-momentum $\boldsymbol{k}$ is defined in terms of $b(y)$ or equivalently $(y,R_\star,\Omega_\star)$, the stress-energy tensor (\ref{eq:radfield}) is completely determined by calculating the quantity $\Phi$, which follows from the conservation equations $\nabla_{\beta}T^{\alpha\beta}=0$. Due to the absence of photon latitudinal motion ($k^{y}=0$) and the axial symmetries of the Erez-Rosen spacetime, we have \cite{DeFalco20183D,Bakala2019}
\begin{equation}\label{flux_cons}
0=\nabla_\beta (\Phi^2 k^\beta)=\partial_x (\sqrt{-g}\Phi^2 k^x).
\end{equation}
Therefore, we obtain 
\begin{equation} \label{eq:phi}
\sqrt{-g}\Phi^2 k^x=\hbox{\rm const} = M(1-y^2)\Phi_0^2,
\end{equation}
where $\Phi_0$ is $\Phi$ evaluated at the emitting surface. Then, after some algebra, we obtain
\begin{equation}\label{INT_PAR}
\Phi^2=\frac{\Phi_0^2\sqrt{(1-y^2)}e^{-\lambda+2\mu}}{M\sqrt{(x^2-y^2)[M^2(x^2-1)(1-y^2)-b^2e^{4\mu}}]}.
\end{equation}
Such formula for $\Phi$ reduces exactly to the Schwarzschild case for $q=0$ \cite{DeFalco20183D,Bakala2019}, and for $\psi=\pi/2$ and $y=0$ to the 2D description in the equatorial plane of Ref. \cite{Bini2015}.

\subsection{Test particle motion}
\label{sec:tpmotion}
A test particle moves in the 3D space with four-velocity $\boldsymbol{U}$ and spatial velocity $\boldsymbol{\hat{\nu}}(\boldsymbol{U},\boldsymbol{n})$ with respect to the local observer frame, given respectively by
\begin{eqnarray} \label{eq:TPvelocity}
\boldsymbol{U}&=&\gamma(\boldsymbol{U},\boldsymbol{n})[\boldsymbol{n}+\boldsymbol{\nu} (\boldsymbol{U},\boldsymbol{n})],\\
\boldsymbol{\hat{\nu}}(\boldsymbol{U},\boldsymbol{n})&=&\nu^{\hat x}\boldsymbol{e_{\hat x}}+\nu^{\hat y}\boldsymbol{e_{\hat y}}+\nu^{\hat \varphi}\boldsymbol{e_{\hat \varphi}}\\
&=&\nu(\sin\alpha\sin\psi\ \boldsymbol{e_{\hat x}}+\cos\psi\ \boldsymbol{e_{\hat y}}+\sin\psi\cos\alpha\ \boldsymbol{e_{\hat \varphi}}),\notag
\end{eqnarray}
where $\gamma(\boldsymbol{U},\boldsymbol{n})\equiv \gamma=1/\sqrt{1-||\boldsymbol{\nu}(\boldsymbol{U},\boldsymbol{n})||^2}$ is the Lorentz factor, $\nu^{\hat\alpha}(\boldsymbol{U},\boldsymbol{n})\equiv \nu^{\hat\alpha}$ is the spatial velocity in the local static observer frame, $\alpha$ and $\psi$ are the azimuthal and polar angle, respectively  \cite{DeFalco20183D,Bakala2019}. The explicit expression of the test particle velocity components are
\begin{equation} \label{eq:velocitycomp}
\begin{aligned}
&U^{\hat t}\equiv \frac{dt}{d\tau}=\frac{\gamma}{e^\mu},\quad U^{\hat x}\equiv \frac{dx}{d\tau}=\frac{\gamma\nu^{\hat x}}{\sqrt{g_{xx}}},\\
&U^{\hat y}\equiv \frac{dy}{d\tau}=\frac{\gamma\nu^{\hat y}}{\sqrt{g_{yy}}},\quad U^{\hat \varphi}\equiv \frac{d\varphi}{d\tau}=\frac{\gamma\nu^{\hat \varphi}}{\sqrt{g_{\varphi\varphi}}},
\end{aligned}
\end{equation}
where $\tau$ is the affine (or proper time) parameter along the test particle's world line.

Using the \emph{observer splitting formalism}, we find the following expression for the test particle acceleration:
\begin{eqnarray}
a(\boldsymbol{U})^{\hat x}&=& \gamma^2 \left[a(\boldsymbol{n})^{\hat x}+\nu^2\left(-k(x,\boldsymbol{n})^{\hat y}\sin\alpha\sin\psi\cos\psi\right.\right.\label{acc1} \\
&&\left.\left.+k(\varphi,\boldsymbol{n})^{\hat x}\sin^2\psi\cos^2\alpha+k(y,\boldsymbol{n})^{\hat x}\cos^2\psi\right)\right]\nonumber\\
&&+\gamma \left(\gamma^2 \sin\alpha\sin\psi \frac{\rm d \nu}{\rm d \tau}+\nu \cos \alpha\sin\psi \frac{\rm d \alpha}{\rm d \tau}\right.\nonumber\\
&&\left.+\nu \cos \psi\sin\alpha \frac{\rm d \psi}{\rm d \tau} \right), \nonumber\\   
a(\boldsymbol{U})^{\hat y}&=&\gamma^2 \left[a(\boldsymbol{n})^{\hat y}+\nu^2\left(k(\varphi,\boldsymbol{n})^{\hat y}\cos^2\alpha\sin^2\psi\right.\right.\label{acc3}\\
&&\left.\left.+k(x,\boldsymbol{n})^{\hat y}\sin^2\alpha\sin^2\psi\right.\right.\nonumber\\
&&\left.\left.-k(y,\boldsymbol{n})^{\hat x}\sin\alpha\sin\psi\cos\psi\right)\right]\nonumber\\
&&+ \gamma\left(\gamma^2 \cos\psi \frac{\rm d \nu}{\rm d \tau}-\nu \sin\psi \frac{\rm d \psi}{\rm d \tau}\right).\nonumber\\
a(\boldsymbol{U})^{\hat \varphi}&=&-\gamma^2 \nu^2\left[k(\varphi,\boldsymbol{n})^{\hat y}\sin\psi\cos\alpha\cos\psi\right.\label{acc2}\\ 
&&\left.+k(\varphi,\boldsymbol{n})^{\hat x}\sin^2\psi\sin\alpha\cos\alpha\right]\nonumber\\
&&+ \gamma\left(\gamma^2 \cos \alpha\sin\psi \frac{\rm d \nu}{\rm d \tau}\right.\nonumber\\
&&\left.-\nu\sin \alpha\sin\psi \frac{\rm d \alpha}{\rm d \tau}+\nu\cos\alpha\cos\psi \frac{\rm d \psi}{\rm d \tau}\right).\nonumber
\end{eqnarray}
From the orthogonality between $\boldsymbol{a}(\boldsymbol{U})$ and $\boldsymbol{U}$, we can determine the expression of $a(\boldsymbol{U})^{\hat t}$ \cite{DeFalco20183D,Bakala2019}
\begin{equation}\label{acc4}
\begin{aligned}
a(\boldsymbol{U})^{\hat t}&=\nu[a(\boldsymbol{U})^{\hat x}\sin\alpha\sin\psi+a(\boldsymbol{U})^{\hat y}\cos\psi\\
&+a(\boldsymbol{U})^{\hat \varphi}\cos\alpha\sin\psi]\\
&=\gamma^2\nu\left[a(\boldsymbol{n})^{\hat x}\sin\alpha\sin\psi+a(\boldsymbol{n})^{\hat y}\cos\psi\right]\\
&+ \gamma^3\nu \frac{\rm d \nu}{\rm d \tau}.
\end{aligned}
\end{equation}

\subsection{Test particle-radiation field interaction}
\label{sec:radfield}
We assume that the radiation-test particle interaction occurs through Thomson scattering, characterized by a constant momentum-transfer cross section $\sigma$, independent from direction and frequency of the radiation field. The radiation force is \cite{Bini2009,Bini2011,DeFalco20183D,Bakala2019}
\begin{equation} \label{radforce}
{\mathcal F}_{\rm (rad)}(\boldsymbol{U})^{\hat\alpha} = -\sigma P(\boldsymbol{U})^{\hat\alpha}{}_{\hat\beta} \, T^{\hat\beta}{}_{\hat\mu} \, U^{\hat\mu} \,,
\end{equation}
where $P(\boldsymbol{U})^{\hat\alpha}{}_{\hat\beta}=\delta^{\hat\alpha}_{\hat\beta}+U^{\hat\alpha} U_{\hat\beta}$ projects a vector orthogonally to $\boldsymbol{U}$. Decomposing the photon four-momentum $\boldsymbol{k}$ first with respect to the test particle four-velocity, $\boldsymbol{U}$, and then in the local observer frame, $\boldsymbol{n}$, we have \cite{DeFalco20183D}
\begin{equation} \label{diff_obg}
\boldsymbol{k} = E(\boldsymbol{n})[\boldsymbol{n}+\boldsymbol{\hat{\nu}}(\boldsymbol{k},\boldsymbol{n})]=E(\boldsymbol{U})[\bold{U}+\boldsymbol{\hat {\mathcal V}}(\boldsymbol{k},\boldsymbol{U})].
\end{equation}
Exploiting Eq. (\ref{diff_obg}) in Eq. (\ref{radforce}), we obtain \cite{DeFalco20183D,Bakala2019}
\begin{equation} \label{Frad0}
\begin{aligned}
{\mathcal F}_{\rm (rad)}(\boldsymbol{U})^{\hat\alpha}&=-\sigma \Phi^2 [P(\boldsymbol{U})^{\hat\alpha}{}_{\hat\beta} k^{\hat\beta}]\, (k_{\hat\mu} U^{\hat\mu})\\
&=\sigma \, [\Phi E(\boldsymbol{U})]^2\, \hat {\mathcal V}(\boldsymbol{k},\boldsymbol{U})^{\hat\alpha}\,.
\end{aligned}
\end{equation}
The equations of motion are $m \bold{a}(\boldsymbol{U}) = \boldsymbol{{\mathcal F}_{\rm (rad)}}(\boldsymbol{U})$, where $m$ is the test particle mass. Defined $\tilde \sigma=\sigma/m$, we obtain the following equations \cite{DeFalco20183D,Bakala2019}
\begin{equation}\label{geom}
\bold{a}(\boldsymbol{U})=\tilde \sigma \Phi^2 E(\boldsymbol{U})^2  \,\boldsymbol{\hat {\mathcal V}}(\boldsymbol{k},\boldsymbol{U}).
\end{equation}
Multiplying scalarly Eq. (\ref{diff_obg}) by $\boldsymbol{U}$, we find \cite{DeFalco20183D,Bakala2019}
\begin{equation} \label{enepart}
E(\boldsymbol{U})=\gamma E(\boldsymbol{n})[1-\nu\sin\psi\cos(\alpha-\beta)].
\end{equation}
Such splitting permits to determine $\boldsymbol{\hat{\mathcal{V}}}(\boldsymbol{k},\boldsymbol{U})=\hat{\mathcal{V}}^t\boldsymbol{n}+\hat{\mathcal{V}}^r\boldsymbol{e_{\hat r}}+\hat{\mathcal{V}}^\theta \boldsymbol{e_{\hat\theta}}+\hat{\mathcal{V}}^\varphi \boldsymbol{e_{\hat\varphi}}$ as \cite{DeFalco20183D,Bakala2019}
\begin{eqnarray}
&&\hat{\mathcal{V}}^{\hat x}=\frac{\sin\beta}{\gamma [1-\nu\sin\psi\cos(\alpha-\beta)]}-\gamma\nu\sin\psi\sin\alpha,\label{rad1}\\
&&\hat{\mathcal{V}}^{\hat y}=-\gamma\nu\cos\psi \label{rad2},\\
&&\hat{\mathcal{V}}^{\hat\varphi}=\frac{\cos\beta}{\gamma [1-\nu\sin\psi\cos(\alpha-\beta)]}-\gamma\nu\sin\psi\cos\alpha,\label{rad3}\\
&&\hat{\mathcal{V}}^{\hat t}=\gamma\nu\left[\frac{\sin\psi\cos(\alpha-\beta)-\nu}{1-\nu\sin\psi\cos(\alpha-\beta)}\right].\label{rad4}
\end{eqnarray}

\subsection{Equations of motion}
\label{sec:eqm}
The test particle equations of motion are (following the same strategy exploited in Refs. \cite{DeFalco20183D,Bakala2019})
\begin{eqnarray}
\frac{d\nu}{d\tau}&=& -\frac{1}{\gamma}\left[a(\boldsymbol{n})^{\hat x}\sin\alpha \sin\psi+a(\boldsymbol{n})^{\hat y}\cos\psi\right]\label{EoM1}\\
&&+\frac{\tilde{\sigma}[\Phi E(\boldsymbol{U})]^2}{\gamma^3\nu}\hat{\mathcal{V}}^{\hat t},\nonumber\\
\frac{d\psi}{d\tau}&=& \frac{\gamma}{\nu} \left\{-a(\boldsymbol{n})^{\hat x}\sin\alpha\cos\psi+a(\boldsymbol{n})^{\hat y}\sin\psi\right.\label{EoM2}\\
&&\left.+\nu^2\left[\left(k(\varphi,\boldsymbol{n})^{\hat y}\cos^2\alpha+k(x,\boldsymbol{n})^{\hat y}\sin^2\alpha\right)\sin\psi\right.\right.\nonumber\\
&&\left.\left.-k(y,\boldsymbol{n})^{\hat x}\sin\alpha\cos\psi\right]\right\}\nonumber\\
&&+\frac{\tilde{\sigma}[\Phi E(\boldsymbol{U})]^2}{\gamma\nu^2\sin\psi}\left[\hat{\mathcal{V}}^{\hat t}\cos\psi-\hat{\mathcal{V}}^{\hat y}\nu\right],\nonumber\\
\frac{d\alpha}{d\tau}&=&\frac{\gamma\cos\alpha}{\nu\sin\psi}\left\{-a(\boldsymbol{n})^{\hat x}\right.\label{EoM3}\\
&&\left.-\nu^2\left[\left(k(\varphi,\boldsymbol{n})^{\hat y}-k(x,\boldsymbol{n})^{\hat y}\right)\sin\psi\cos\psi\sin\alpha\right.\right.\nonumber\\
&&\left.\left.+k(\varphi,\boldsymbol{n})^{\hat x}\sin^2\psi+k(y,\boldsymbol{n})^{\hat x}\cos^2\psi\right]\right\}\nonumber\\
&&+\frac{\tilde{\sigma}[\Phi E(\boldsymbol{U})]^2\cos\alpha}{\gamma\nu\sin\psi}\left[\hat{\mathcal{V}}^{\hat x}-\hat{\mathcal{V}}^{\hat \varphi}\tan\alpha\right],\nonumber\\
U^x&\equiv&\frac{dx}{d\tau}=\frac{\gamma\nu\sin\alpha\sin\psi}{\sqrt{g_{xx}}}, \label{EoM4}\\
U^y&\equiv&\frac{dy}{d\tau}=\frac{\gamma\nu\cos\psi}{\sqrt{g_{yy}}} \label{EoM5},\\
U^\varphi&\equiv&\frac{d\varphi}{d\tau}=\frac{\gamma\nu\cos\alpha\sin\psi}{\sqrt{g_{\varphi\varphi}}}.\label{EoM6}
\end{eqnarray}

Defining $A=\tilde{\sigma}\Phi_0^2E^2$, which is the so-called luminosity parameter, which can be also defined as $A/M=L/L_{\rm EDD}\in[0,1]$, where $L$ is the luminosity measured by a static observer at infinity, and $L_{\rm EDD}$ is the Eddington luminosity \cite{DeFalco20183D,Bakala2019}. Using Eqs. (\ref{INT_PAR}) and (\ref{enepart}), we obtain
\begin{equation}
\tilde{\sigma}[\Phi E(\boldsymbol{U})]^2=\frac{A\gamma^2\sqrt{1-y^2}e^{-\lambda}[1-\nu\sin\psi\cos(\alpha-\beta)]^2}{M \sqrt{(x^2-y^2)[M^2(x^2-1)(1-y^2)-b^2e^{4\mu}}}.
\end{equation}
We note that Eqs. (\ref{EoM1}) -- (\ref{EoM6}) reduce to the Schwarzschild case for $q\to0$ \cite{DeFalco20183D,Bakala2019}, and to the 2D equatorial plane case for $\psi\to\pi/2$ and $y=0$ \cite{Bini2015}. In Appendix \ref{sec:WFL}, we calculate the weak field approximation of Eqs. (\ref{EoM1}) -- (\ref{EoM6}), respectively.

\section{Critical hypersurfaces}
\label{sec:CH}
The dynamical system governed by Eqs. (\ref{EoM1}) -- (\ref{EoM6}) admits a critical hypersurface outside the emitting surface, where the gravitational attraction, and the radiation pressure balance \cite{Bini2015,DeFalco20183D,Bakala2019}. Such region is analytically determined by the critical radius $x_{\rm crit}$ as function of $y$, i.e., $x_{\rm crit}=x_{\rm crit}(y)$, once the parameters $(q,A,R_\star,\Omega_\star)$ are assigned. We consider a test particle moving along a non-equatorial plane on purely circular orbit (i.e., $\alpha=0,\pi$, $\psi=\pi/2$, and $\nu=\mbox{const}$). Equation (\ref{EoM1})
for $d\nu/d\tau=0$ reduces to \cite{Bini2015,DeFalco20183D,Bakala2019}
\begin{equation} 
\tilde{\sigma}[\Phi E(\boldsymbol{U})]^2\hat{\mathcal{V}}^{\hat t}=0,\quad \Rightarrow\quad \nu=\cos\beta.
\end{equation}
The velocity of the test particle equates the photon azimuthal velocity (see Ref. \cite{DeFalco20183D,Bakala2019}, for further details). Since the test particle moves tangentially on the critical hypersurface, we have $d\alpha/d\tau=0$, and Eq. (\ref{EoM3}) becomes
\begin{equation} \label{eq:CH}
a(\boldsymbol{n})^{\hat x}-k(\varphi,\boldsymbol{n})^{\hat x}\nu^2=\frac{\tilde{\sigma}[\Phi E(\boldsymbol{U})]^2}{\gamma^2}\hat{\mathcal{V}}^{\hat x},
\end{equation}
which is the implicit equation defining the critical hypersurface. Naturally, Eq. (\ref{eq:CH}) reduces to the Schwarzschild case for $q=0$ \cite{DeFalco20183D,Bakala2019}. The critical hypersurface is axially symmetric  with respect to the polar direction \cite{DeFalco20183D,Bakala2019}, and can assume either an oblate or prolate form.

In the next sections, we analyse in detail the proprieties of the critical hypersurfaces (Sec. \ref{sec:PCH}), we show some selected test particle orbits (Sec. \ref{sec:TPO}), we investigate the condition for the multiplicity of critical hypersurface solutions (Sec. \ref{sec:MCH}), we perform the calculations to obtain suspended orbits (Sec. \ref{sec:SO}), and finally we graphically show the stability of the critical hypersurfaces through the Lyapunov functions (Sec. \ref{sec:SCH}).

\subsection{Proprieties of critical hypersurfaces}
\label{sec:PCH}
\begin{figure}[htbp]  
\begin{center}
\vbox{
\hbox{
\includegraphics[scale=0.15]{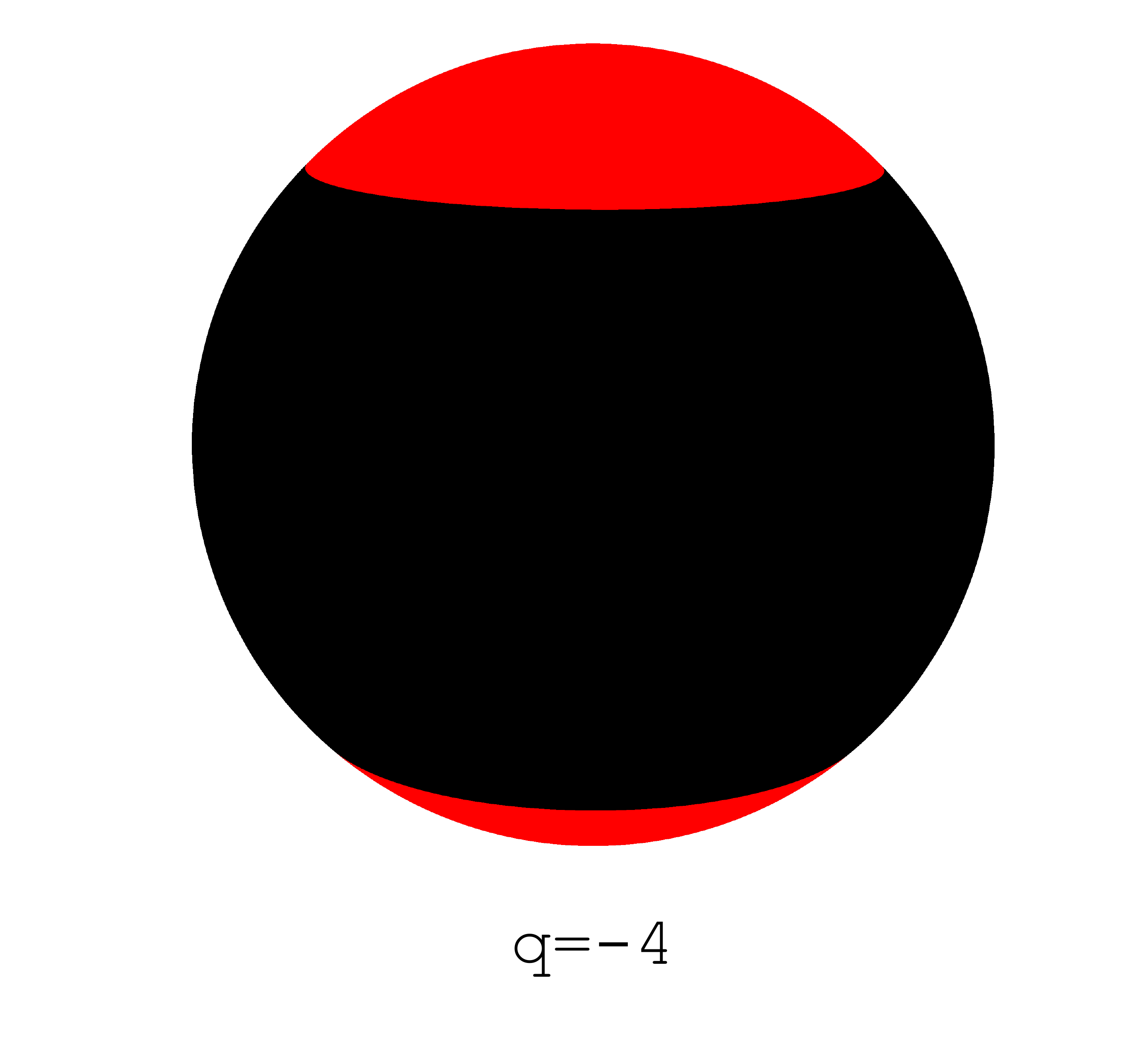}
\includegraphics[scale=0.15]{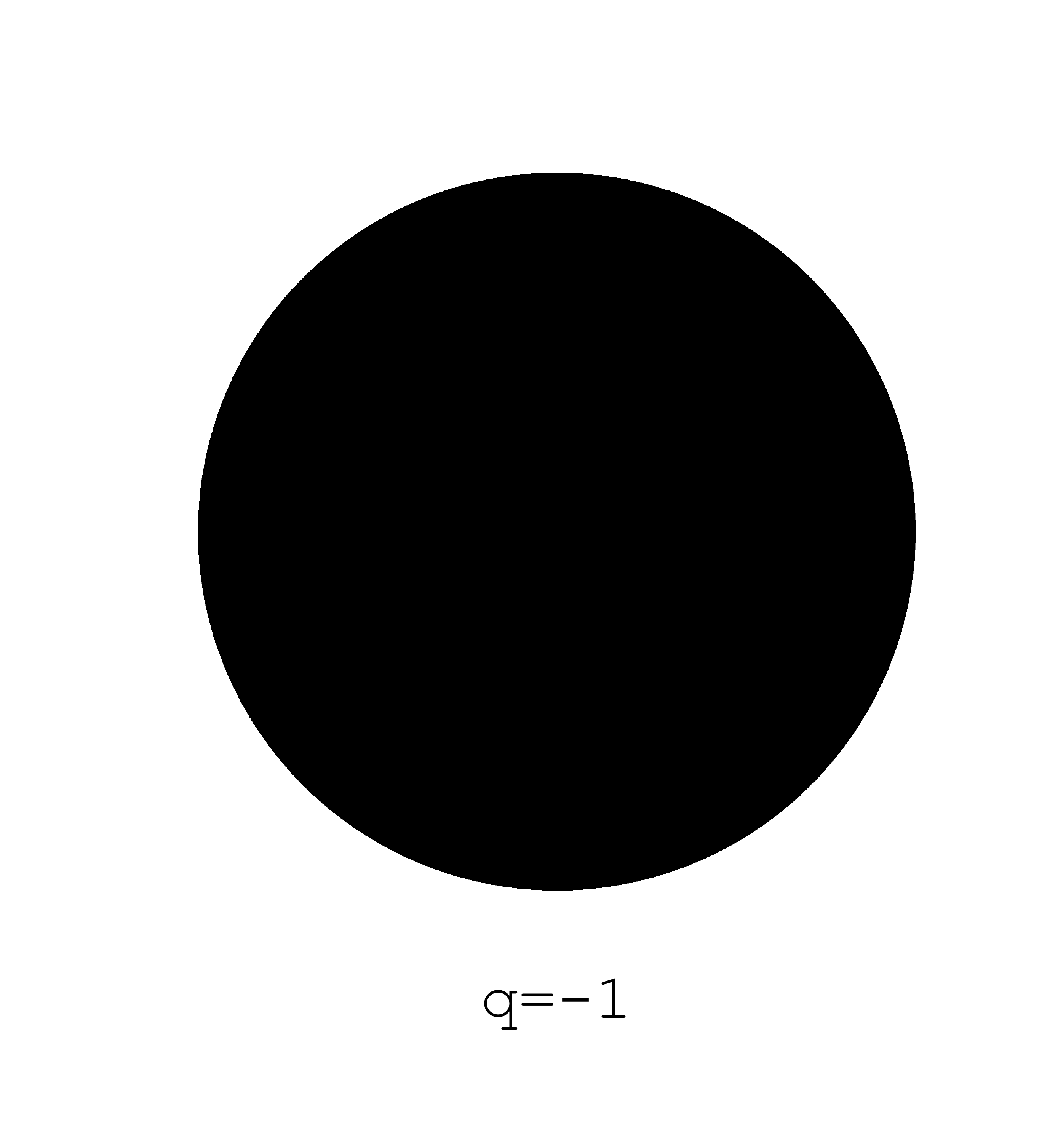}
}
\hbox{
\includegraphics[scale=0.15]{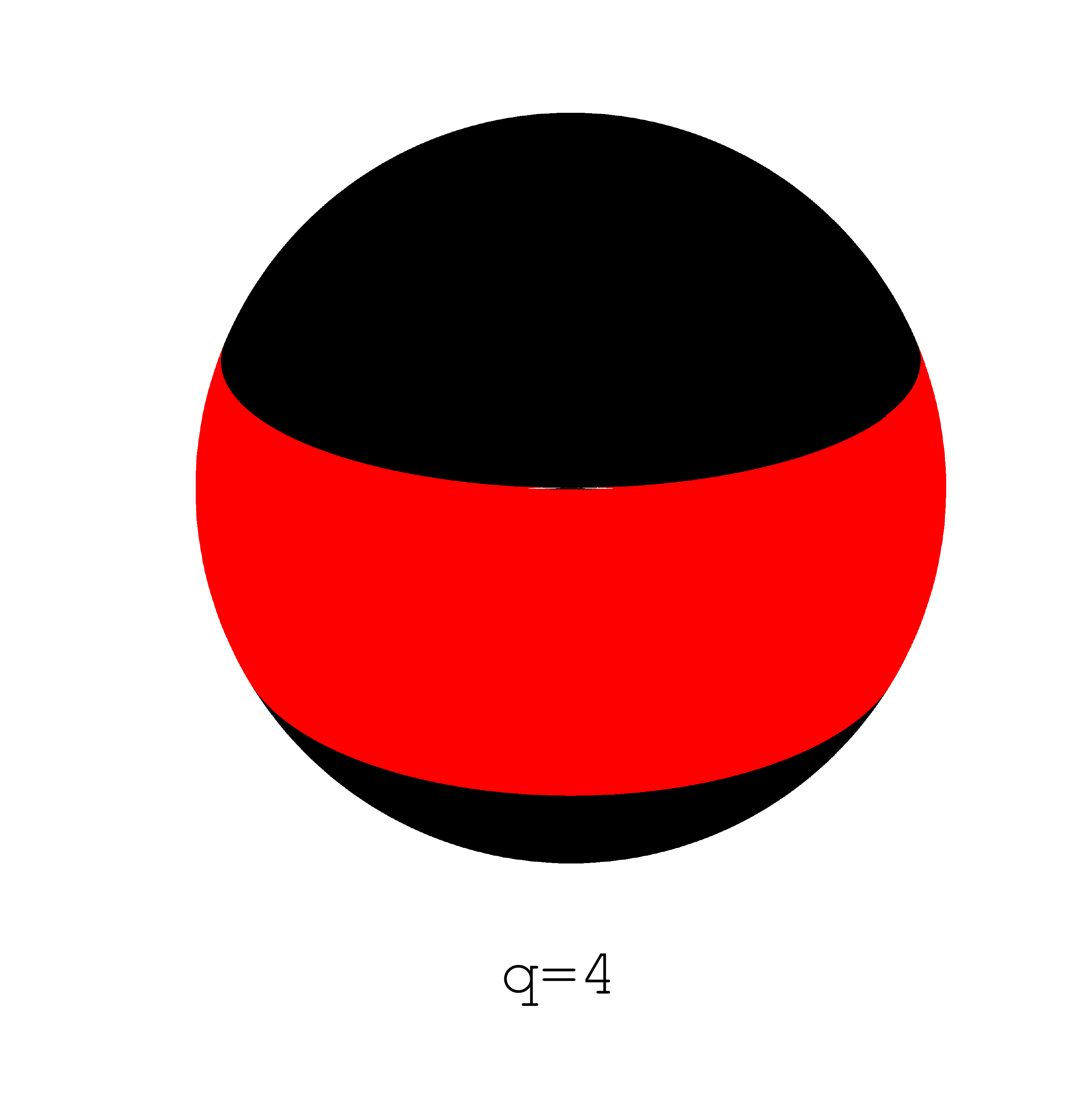}
\includegraphics[scale=0.15]{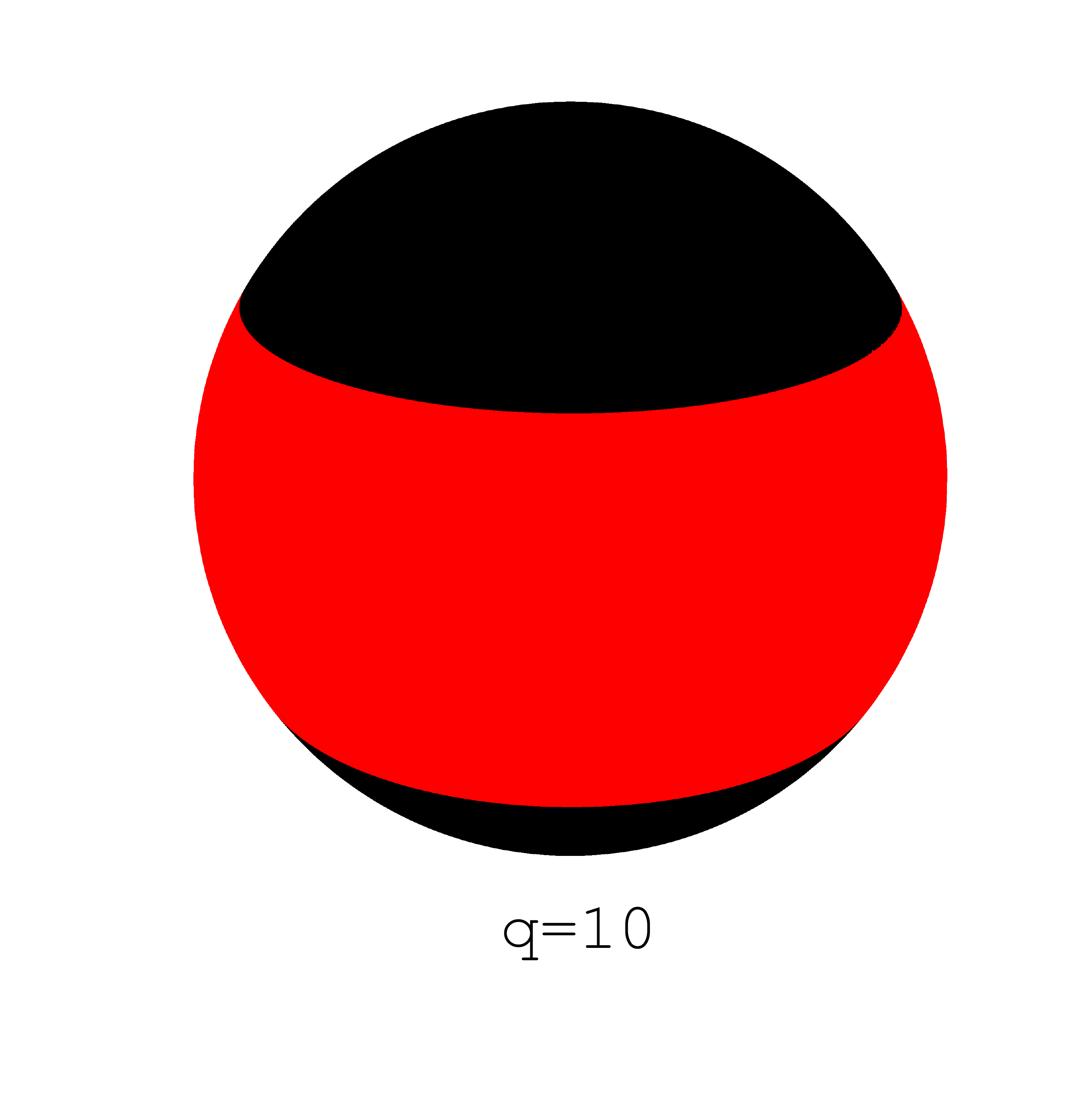}
}}
\end{center}
\caption{Different configurations of event horizon (black region) and Killing hypersurface (red region) in the Erez-Rosen spacetime for $q=-4,-1,4,10$.}  
\label{fig:Fig1}
\end{figure}

In this and next sections, we focus our attention only on BH cases (with $M=5M_\odot$), even though all what we develop can be also easily adapted to the NS case. It is important to first analyse some useful proprieties of the event horizon hypersurface. This is a null region, where the redshift function, with respect to a static observer located at infinity, becomes infinite and it is characterized by the condition $g_{tt}=0$ \cite{Quevedo1990}, which implies $x=1$, and 
\begin{equation} 
\begin{cases}
\mbox{for}\ q>0 & \frac{1}{3}\left(1-\frac{2}{q}\right)\le y^2\le 1,\\
\mbox{for}\ q=0 & \mbox{Schwarzschild\ case},\\
\mbox{for}\ q<0 & 0\le y^2\le \frac{1}{3}\left(1-\frac{2}{q}\right).
\end{cases}
\end{equation}
It is important to note that only for $-1\le q\le2$ the horizon totally covers the hypersurface $x=1$, while in the other cases the parts that are not covered by the horizon are defined \emph{Killing singularities} (i.e., hypersurfaces where the norm of the timelike Killing vector becomes infinite) \cite{Quevedo1990}. In Fig. \ref{fig:Fig1}, we show some graphical examples of event horizon and Killing singularity for different values of $q$ (see Ref. \cite{Quevedo1990}, for more details). We report some examples of the event horizon hypersurface, and do not analyse their proprieties or further implication in the 3D space, because this goes beyond the aim of our paper. We reserve such geometrical and physical investigations in a future paper, since, at the best of our knowledge, they have never been considered in the literature.

\begin{figure}[htbp]  
\begin{center}
\includegraphics[scale=0.3]{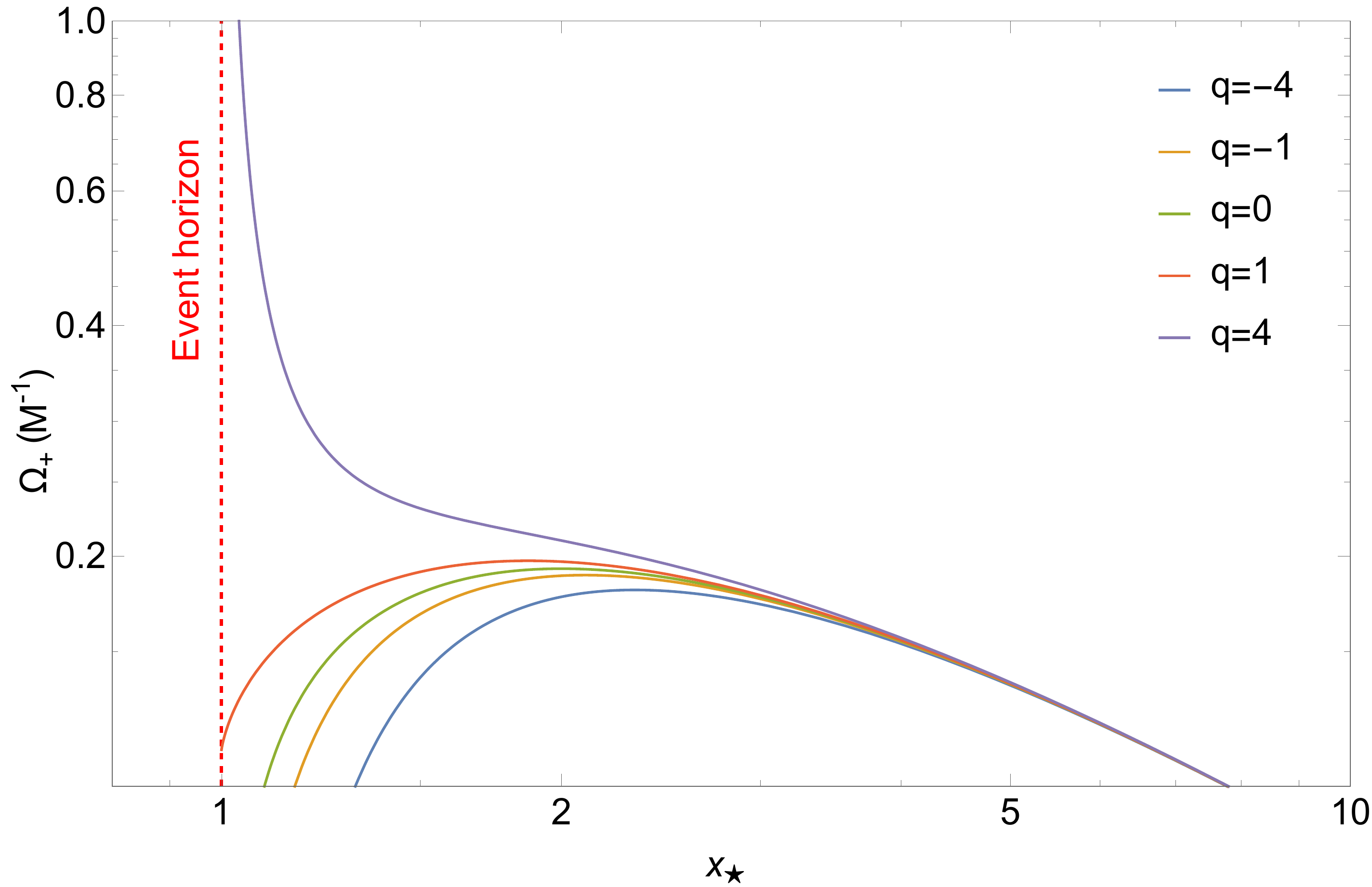}
\end{center}
\caption{Maximum angular velocity $\Omega_+$ in terms of the emitting source radius $x_\star$ for different values of the quadrupole moment $q=-4,-1,0,1,4$. The dashed red line represents the BH event horizon in the Erez-Rosen metric.}  
\label{fig:Fig2}
\end{figure}

\begin{figure}[htbp]  
\begin{center}
\includegraphics[trim=3cm 0cm 0cm 0cm,scale=0.39]{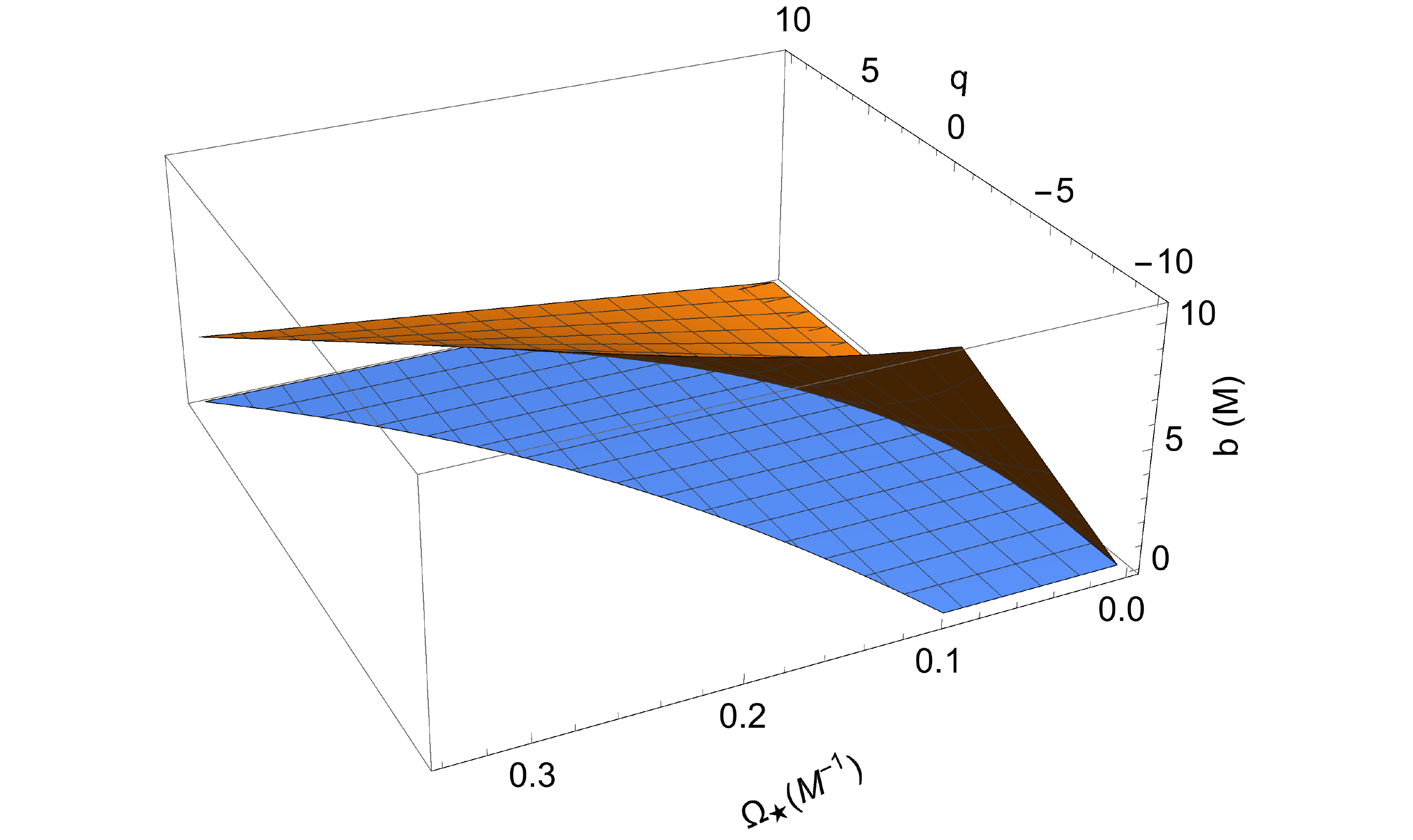}
\end{center}
\caption{Azimuthal photon impact parameter plotted in terms of $q$ and $\Omega_\star\in[0,\Omega_+(x_\star,0,q)]$ with $x_\star=1.5$. The orange and blue surfaces are plotted for $y=0$ and $y=1$, respectively.}  
\label{fig:Fig3}
\end{figure}

The angular velocity of the emitting surface $\Omega_\star$ is bounded by the upper and lower values $\Omega(x_\star,y=0,q)_\pm=\pm\sqrt{-g_{tt}/g_{\varphi\varphi}}$ (see Refs. \cite{Bakala2019}, for details). Since we consider $\Omega_\star\ge0$, and $\Omega_-=-\Omega_+\le0$, we have
\begin{equation}
0\le \Omega_\star\le\Omega_+, 
\end{equation}
otherwise beyond such boundaries we obtain unphysical superluminal rotations. In Fig. \ref{fig:Fig2}, we show how $\Omega_+$ changes in terms of the emitting source radius $x_\star$, and different values of $q$.
We note that for $q<-1$, we have $\Omega_+\to\infty$; while for $q\ge-1$, $\Omega_+$ is upper bounded. The maximum extension of the emitting surface locates around $x_\star=7.73$, instead the minimum depends on the value of $q$ and can even be very close to $x_\star=1$. Such constraints impose also limits on the $b$-range, as it can be seen in  Fig. \ref{fig:Fig3}, where we display the photon azimuthal impact parameter $b$ as function of $q$ and $\Omega_\star$.

\begin{figure*}[htbp]  
\begin{center}
\hbox{
\includegraphics[scale=0.32]{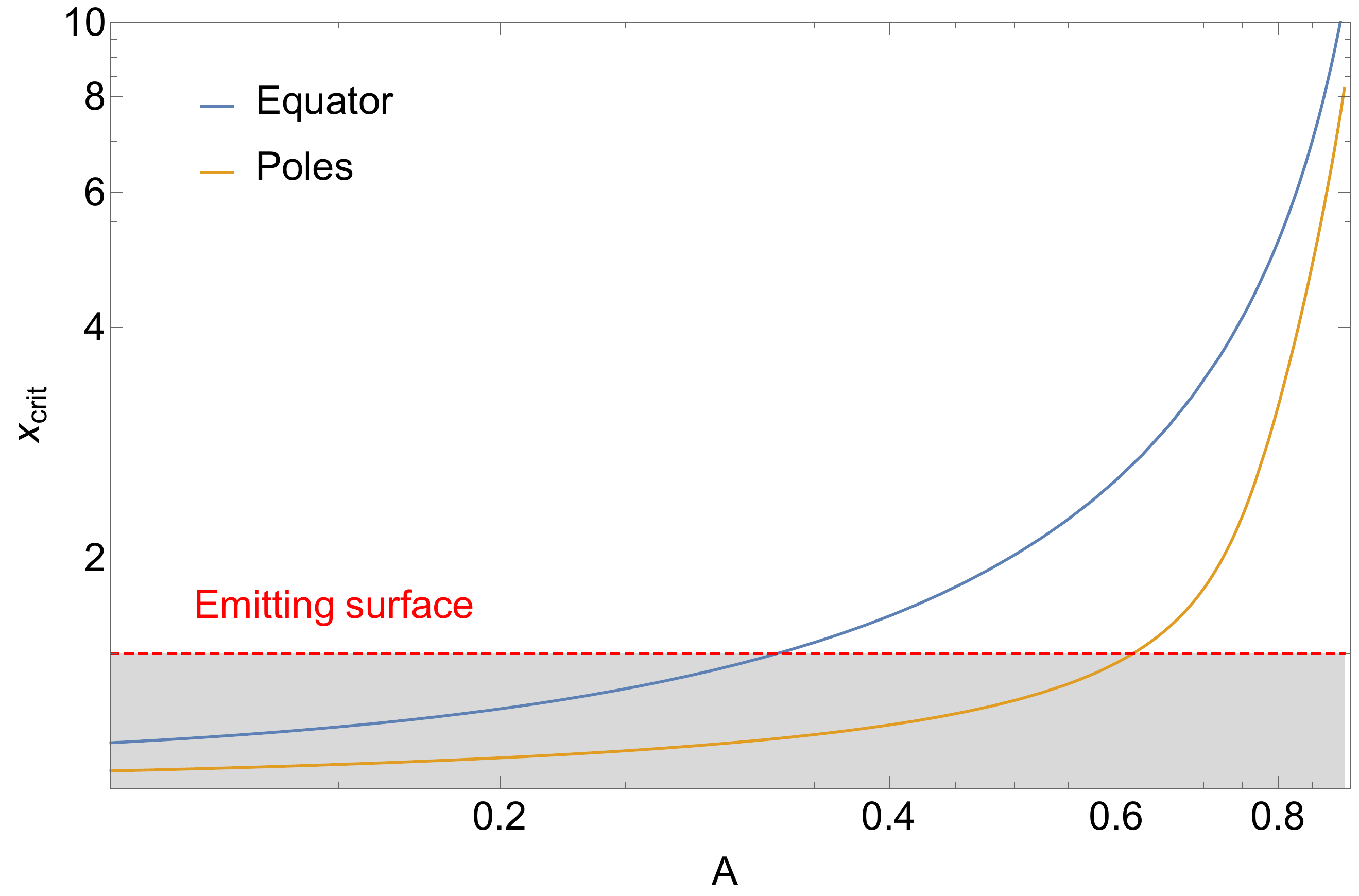}
\hspace{0.3cm}
\includegraphics[scale=0.32]{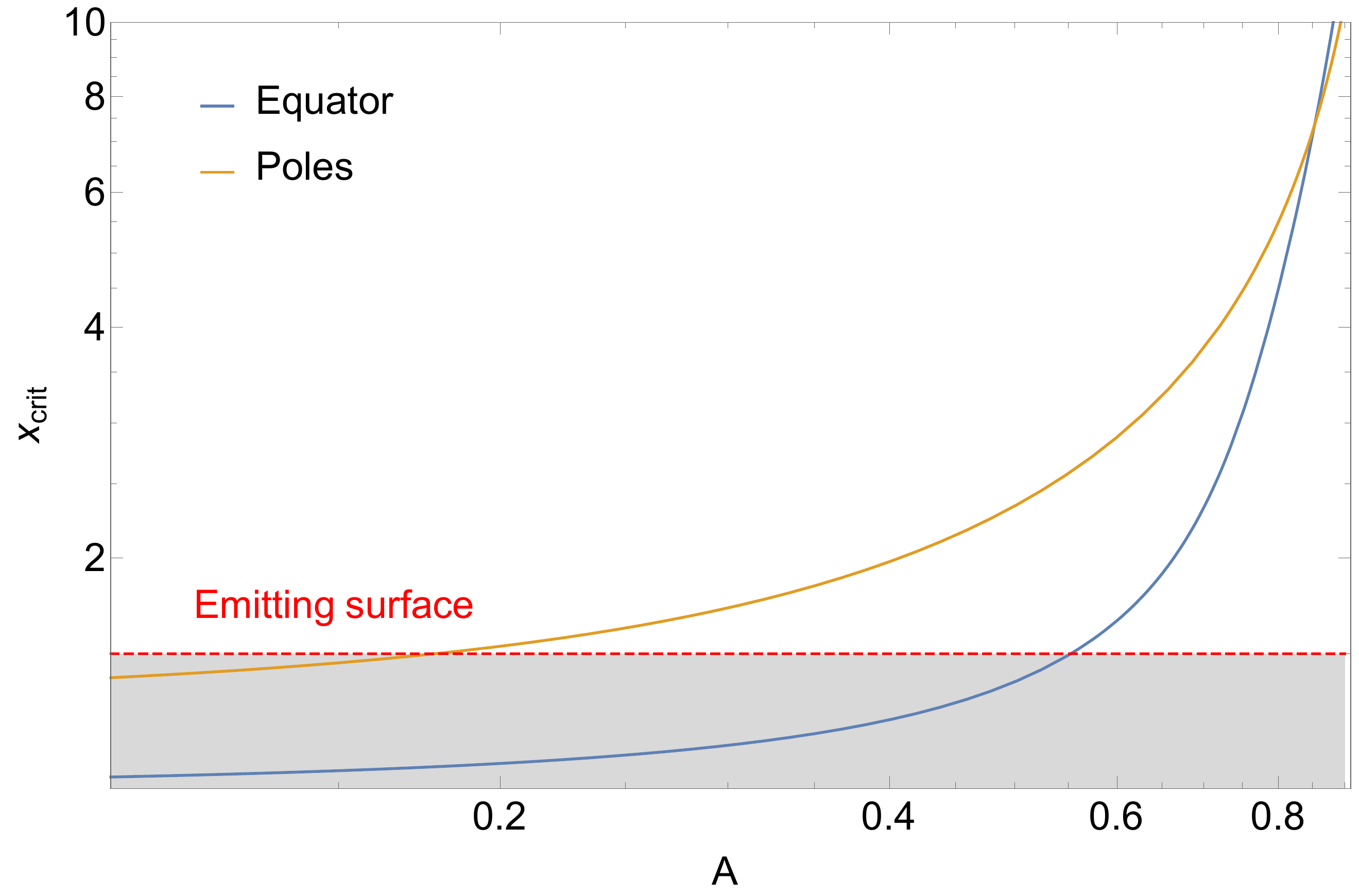}
}
\end{center}
\caption{Critical radius plotted in terms of the luminosity parameter $A$ at the equator (blue line) and at the poles (orange line), having set  
$x_\star=1.5$ (dashed red line), and $\Omega_\star=0.017$ for both plots and $q=3$ for the left panel and $q=-3$ for the right panel. The gray area represents the unphysical solutions.}  
\label{fig:Fig4}
\end{figure*}
Now, we analyse the critical hypersurface behaviors in terms of the parameters $A,\Omega_\star$, distinguishing the cases for negative and positive quadrupole moment $q$. 

In Fig. \ref{fig:Fig4} we report the behavior of the equatorial and polar critical radii as functions of the luminosity parameter $A$ for $q=\pm3$, $x_\star=1.5$, and $\Omega_\star=0.017$. The shape of the critical hypersurface changes in terms of the value of the quadrupole moment, deducing that for $q>2$ it becomes oblate, while for $q<-1$ is prolate. In addition, another change of form occurs at high luminosity for $A\sim0.8$. This plot finds a physical explanation on the mass distribution concentred toward the polar axis ($q>2$), where there is a stronger gravitational pull, allowing thus an extension of the critical hypersurface in the radial direction; viceversa ($q<-1$), employing the same argument for mass distribution towards the radial direction, it entails a prolate shape. All the solutions lying under the emitting surface radius $x_\star$ are considered unphysical throughout the paper, because the test particle cannot penetrate this surface (see \cite{Bakala2019}, for details).

\begin{figure*}[htbp]  
\begin{center}
\hbox{
\includegraphics[scale=0.32]{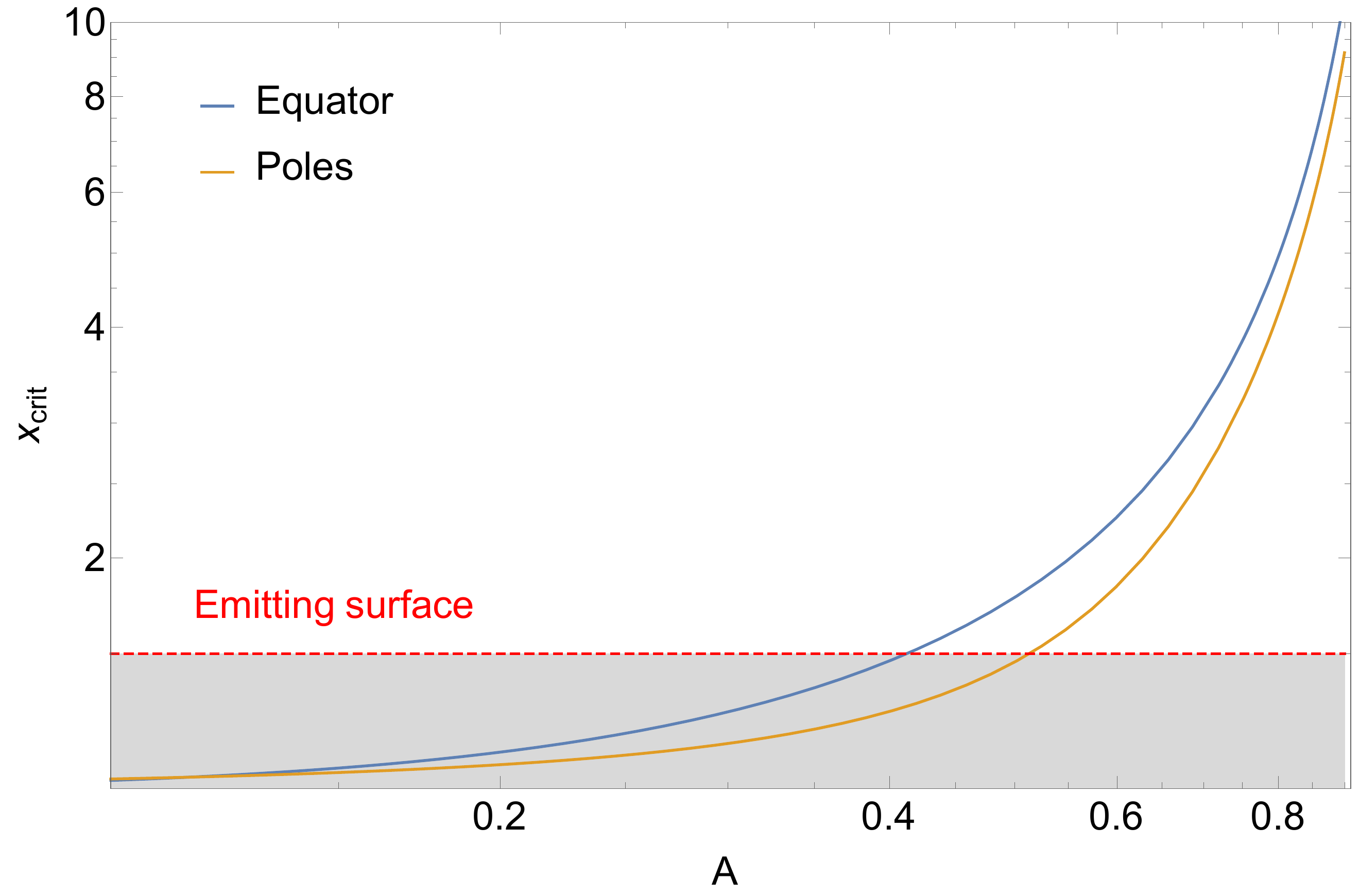}
\hspace{0.3cm}
\includegraphics[scale=0.32]{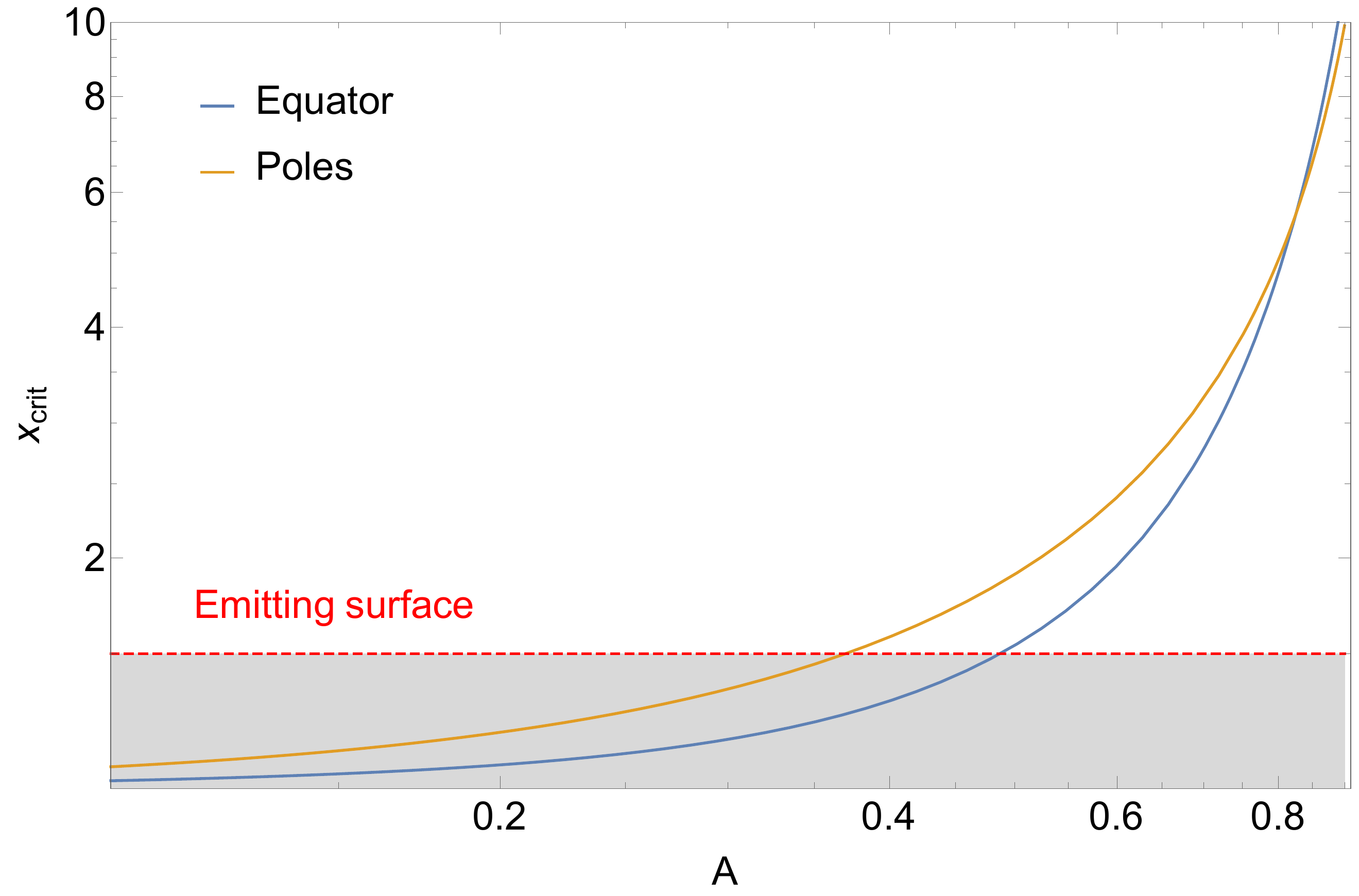}
}
\end{center}
\caption{Critical radius plotted in terms of the luminosity parameter $A$ at the equator (blue line) and at the poles (orange line), having set  
$x_\star=1.5$ (dashed red line), and $\Omega_\star=0.051$ for both plots and $q=1$ for the left panel and $q=-1$ for the right panel. The gray area represents the unphysical solutions.}  
\label{fig:Fig5}
\end{figure*}
A special treatment must be reserved to the interval $-1\le q\le2$, where the metric assumes a particular behavior. In Fig. \ref{fig:Fig5} we plot the polar and equatorial critical radii for $x_\star=1.5$, $\Omega_\star=0.051$, and $q=\pm1$. Both cases show the same trend, but for $q=-1$ (left panel) we have an inversion of shape at $A\sim0.7$. Instead, for $q=1$ (right panel) there is a change of shape at $A\sim0.13$, but since it occurs under the emitting surface it is not taken into account. We have checked that such behavior remains still true also for different angular velocities $\Omega_\star$. Therefore, we conclude that a change of form can occur only in the range $q\in[-\infty,2]$ and for high luminosity $A\gtrsim0.7$. Instead, for $q=\pm1$ we are close to a spherical mass distribution, therefore the critical hypersurfaces assume the limiting behaviors of the Schwarzschild metric.
\begin{figure}[htbp]  
\begin{center}
\includegraphics[scale=0.3]{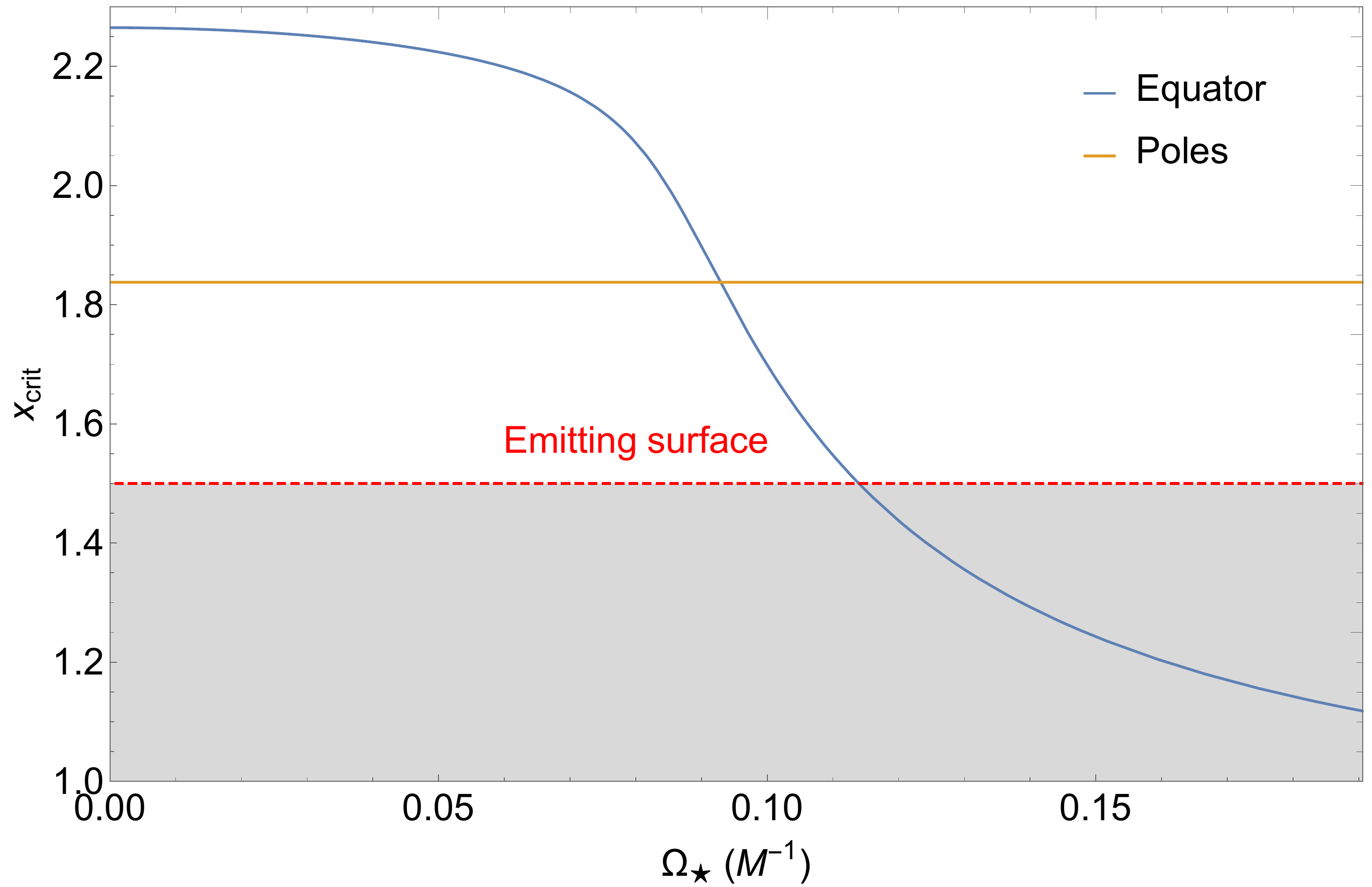}
\end{center}
\caption{Critical radius at the equator (blue line) and at the poles (orange line) in terms of the emitting surface's angular velocity $\Omega_\star$, set $q=1$, $A=0.6$, and $x_\star=1.5$ (dashed red line). The gray area represents the unphysical solutions.}  
\label{fig:Fig6}
\end{figure}
Instead, if we consider the critical radius in terms of the emitting surface's angular velocity $\Omega_\star$ we note that for $q<0$ we have always a prolate shape for the whole range of luminosities, while for $q>2$ always an oblate form. In Fig. \ref{fig:Fig6} we show the case for $q=1$. In such particular situation, there is the intersection between the critical radius at poles and equator occurring at $\Omega_\star=0.092$, which in turn depends on the luminosity parameter $A$. We note that the higher is $A$, the more the intersection occurs at slower and slower angular velocities.  

\begin{figure}[htbp]  
\begin{center}
\includegraphics[scale=0.33]{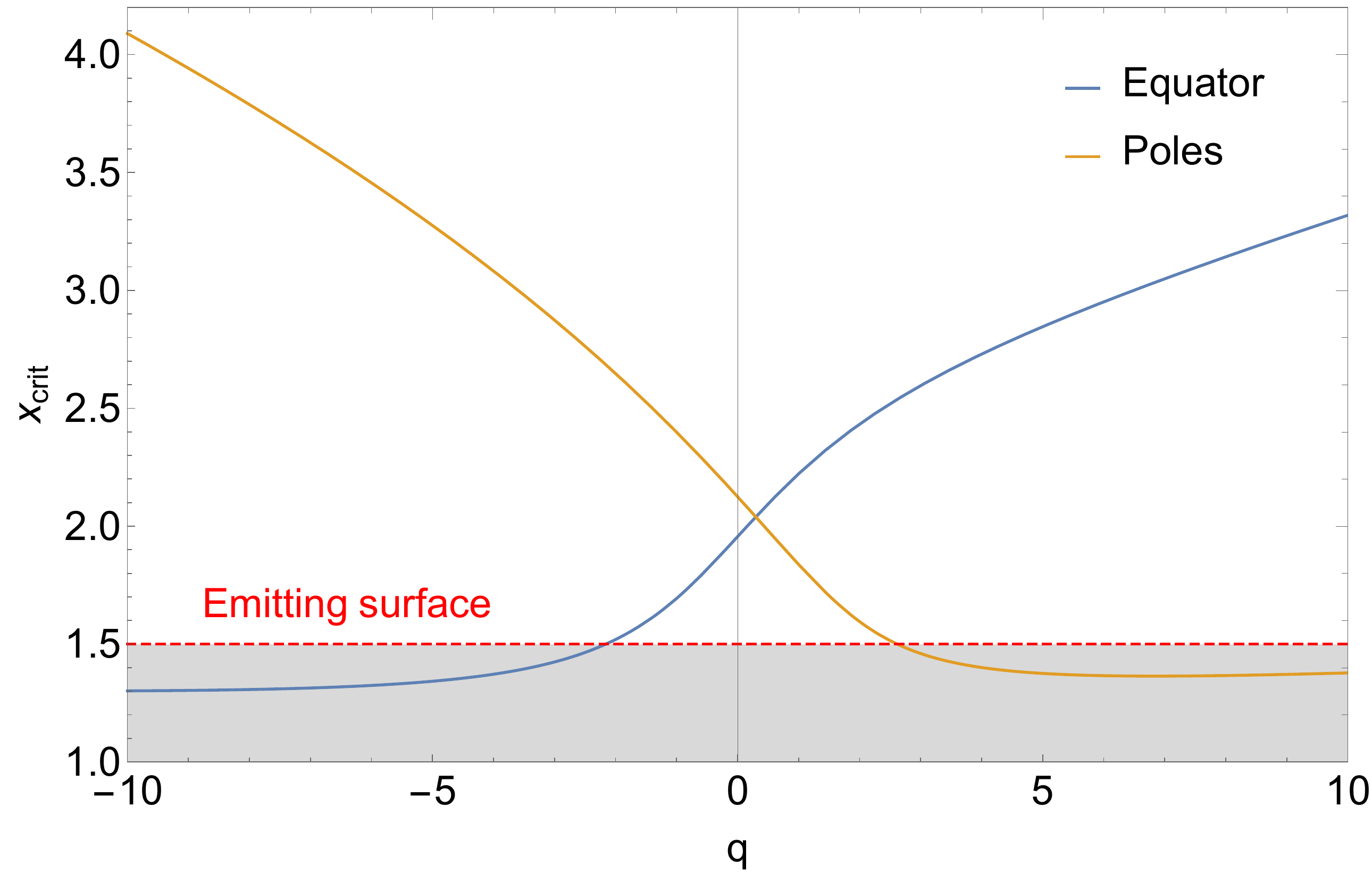}
\end{center}
\caption{Critical radius at the equator (blue line) and at the poles (orange line) in terms of the quadrupole moment $q$, set $A=0.6$, $\Omega_\star=0.017$, and $x_\star=1.5$ (dashed red line). The gray area represents the unphysical solutions.}  
\label{fig:Fig7}
\end{figure}
As a summary of what we have deduced so far,  we report in Fig. \ref{fig:Fig7} the critical radius in terms of the $q$ parameter. This plot is consistent with our results, because for $q<0$ we have a prolate shape, while for $q>0$ the critical hypersurface assumes an oblate shape. In particular the intersection point, where we have an inversion of shape ranges in the interval $0\le q\le 1$. In addition, the higher the luminosities and angular velocities are, the more the intersection point locates towards $q=1$, because the emitting surface breaks the spherical symmetry of the Schwarzschild metric ($q=0$).

\begin{figure}[htbp]  
\begin{center}
\includegraphics[scale=0.31]{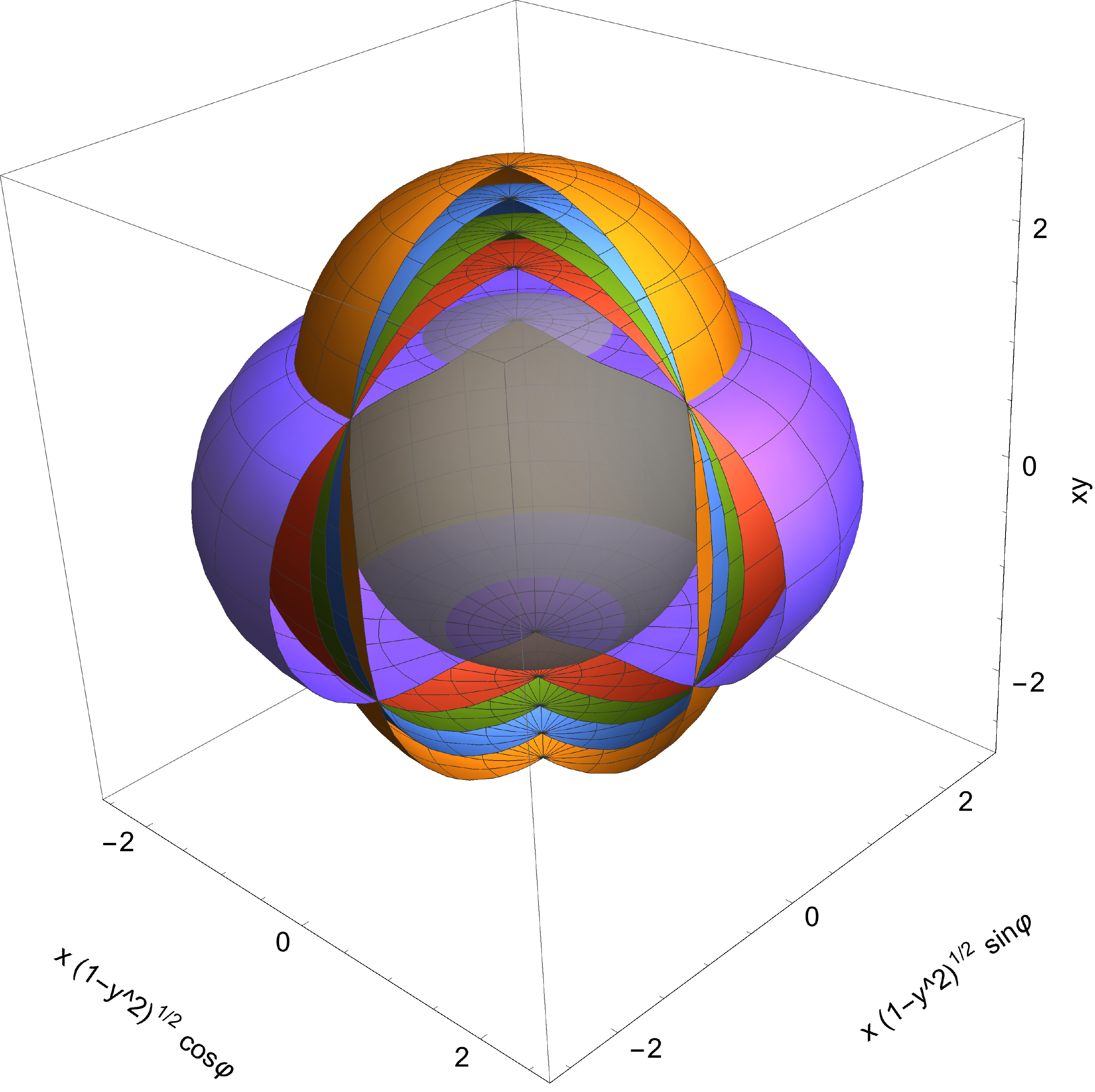}
\end{center}
\caption{Critical hypersurfaces for $q=-2$ (blue region), $q=-1$ (orange region), $q=0$ (green region), $q=1$ (red region), $q=4$ (violet region)
set $A=0.6$, $\Omega_\star=0.017$, and $x_\star=1.5$. The gray spherical region represents the emitting surface. The respective critical radii at the equator and poles are $x_{\rm crit}^{\rm eq}\sim1.80,1.95,2.11,2.26,2.65$, and $x_{\rm crit}^{\rm pole}\sim2.65,2.40,2.13,1.84,1.40$, respectively.}  
\label{fig:Fig8}
\end{figure}
Finally, in Fig. \ref{fig:Fig8} we display some 3D critical hypersurfaces for different values of $q$. We note that they are all very close to the compact object, even for high luminosities. Such configurations can be fundamental for testing Einstein theory in strong field regimes. 
Increasing the values of $q>0$, we have intersection with the emitting surface toward the poles; instead decreasing $q<0$, we have intersection toward the equator. 

\subsection{Test particle orbits}
\label{sec:TPO}
\begin{figure*}
\begin{center}
\vbox{
\hbox{
\includegraphics[scale=0.31]{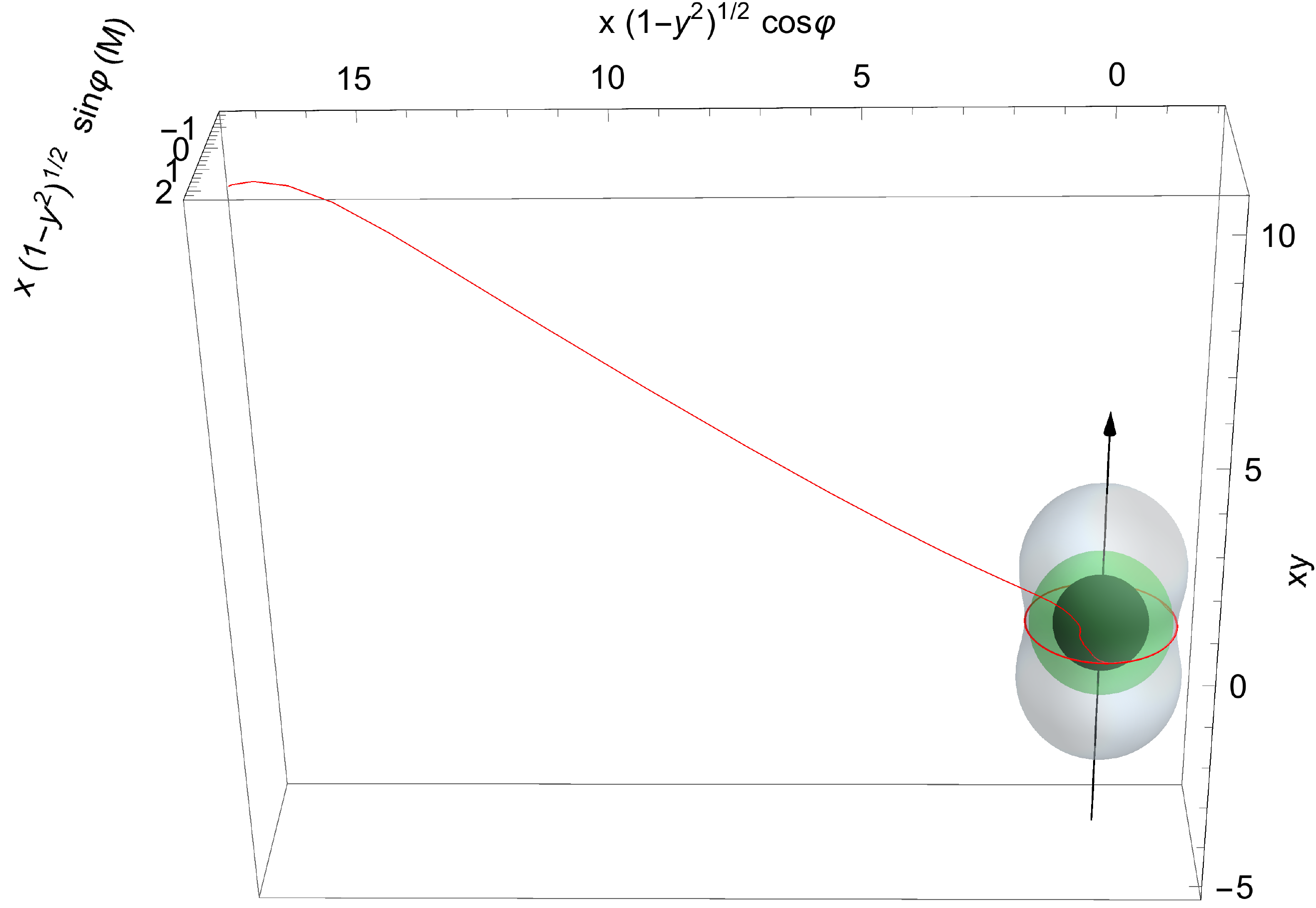}
\hspace{0.1cm}
\includegraphics[scale=0.33]{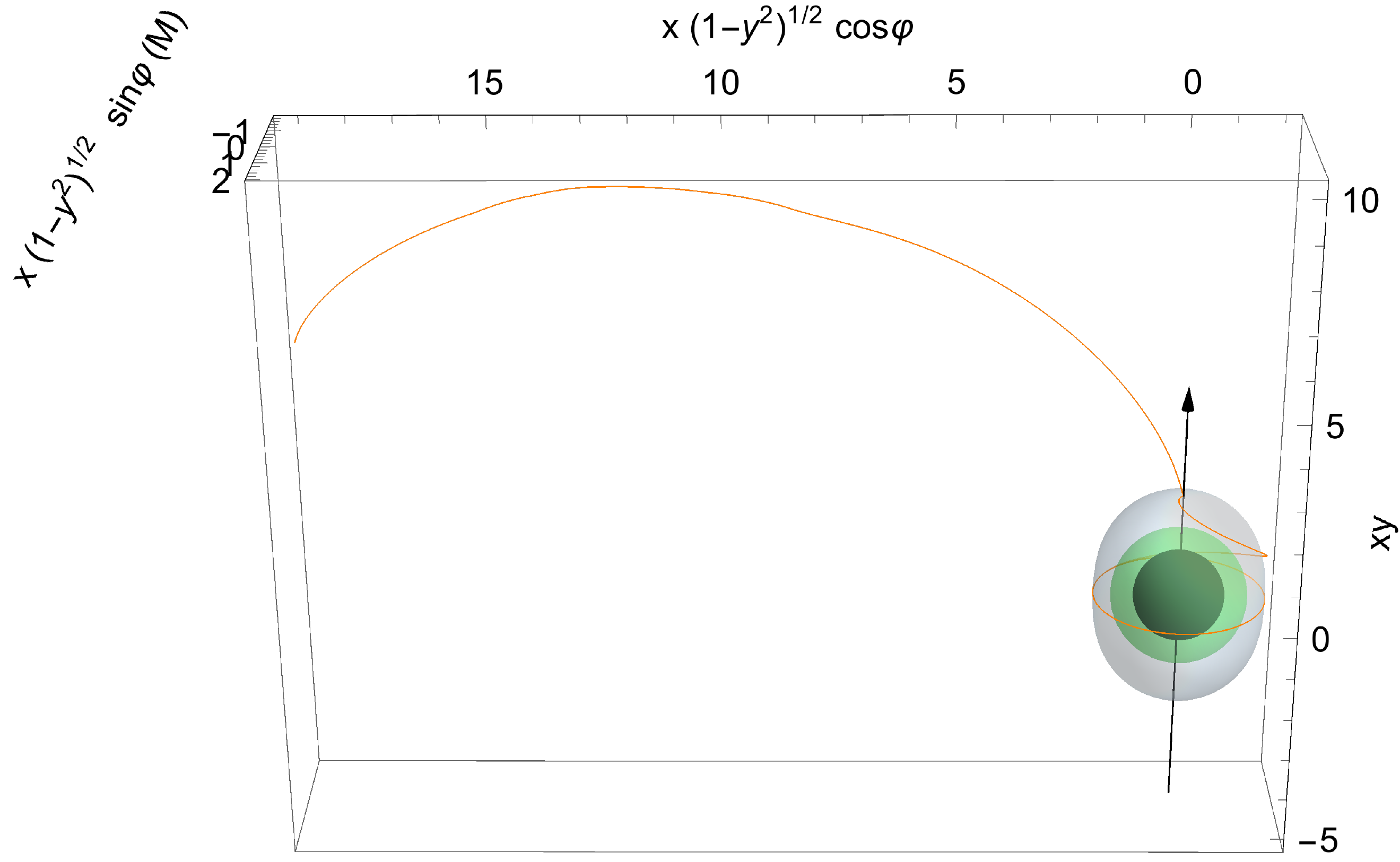}}
\vspace{0.1cm}
\hbox{\hspace{0.5cm}
\includegraphics[scale=0.25]{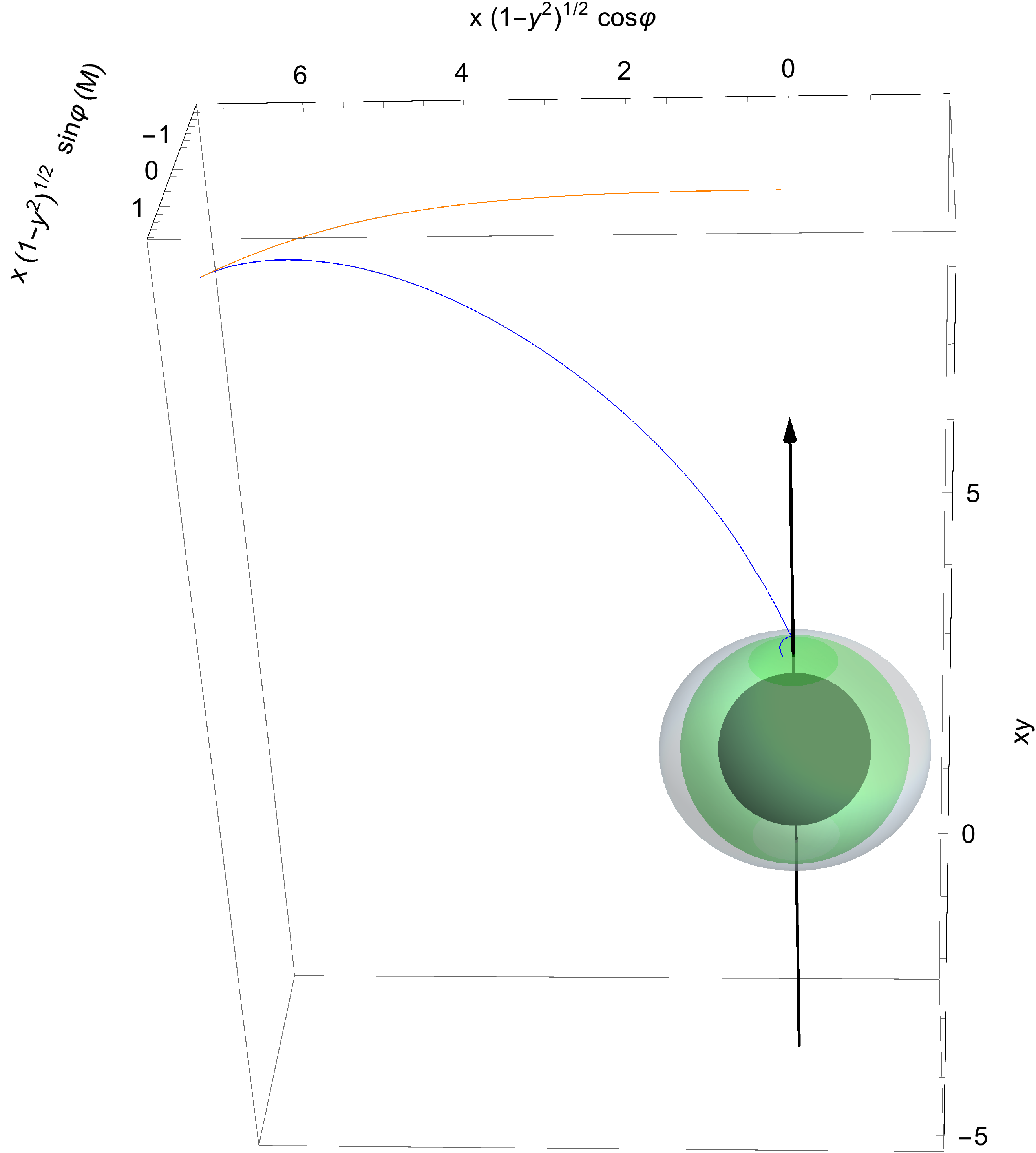}
\hspace{2.5cm}
\includegraphics[scale=0.3]{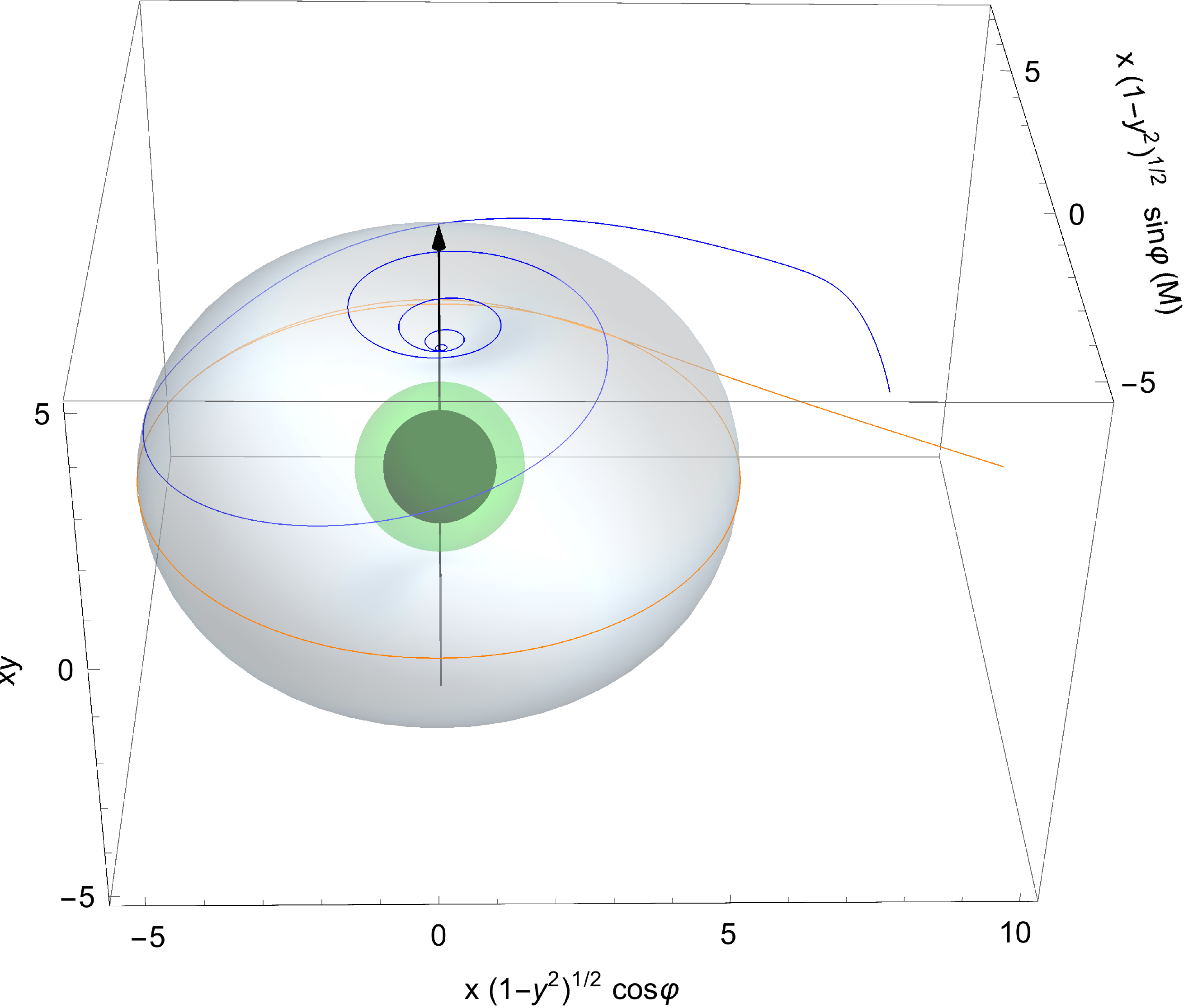}}}
\end{center}
\caption{Examples of test particle trajectories interacting with the radiation field emitted by a spherical and rigidly rotating radiation surface outside of a static and non-spherical BH. 
Top left panel: The case of quadrupole moment $q=-4$, luminosity parameter $A=0.6$, and radiation source angular velocity $\Omega_\star=7.74\times10^{-3}\ M^{-1}$. The test particle starts at the initial position $(x_0,y_0,\phi_0)=(20,0.5,0)$ with initial velocity $(\nu_0,\psi_0,\alpha_0)=(0.3,\pi/3,0)$.
Top right panel: The case of quadrupole moment $q=-1$, luminosity parameter $A=0.6$, and radiation source angular velocity $\Omega_\star=30.94\times10^{-3}\ M^{-1}$. The test particle starts at the initial position $(x_0,y_0,\phi_0)=(20,0.3,0)$ with initial velocity $(\nu_0,\psi_0,\alpha_0)=(0.1,\pi/4,0)$.
Bottom left panel: The case of quadrupole moment $q=1$, luminosity parameter $A=0.5$, and radiation source angular velocity $\Omega_\star=12.38\times10^{-3}\ M^{-1}$. One test particle (blue line) is emitted at the initial position $(x_0,y_0,\phi_0)=(10,0.7,0)$ with initial velocity $(\nu_0,\psi_0,\alpha_0)=(0.1,\pi/10,-\pi/3)$, while the other (orange line) starts at $(x_0,y_0,\phi_0)=(10,0.7,0)$ with initial velocity $(\nu_0,\psi_0,\alpha_0)=(0.4,\pi/10,-\pi/3)$.
Bottom right panel: The case of quadrupole moment $q=4$, luminosity parameter $A=0.8$, and radiation source angular velocity $\Omega_\star=15.47\times10^{-3}\ M^{-1}$. One test particle (blue line) starts at the initial position $(x_0,y_0,\phi_0)=(8,0.2,0)$ with initial velocity $(\nu_0,\psi_0,\alpha_0)=(0.2,\pi/2,0)$, while the other (orange line) is emitted at $(x_0,y_0,\phi_0)=(10,0,0)$ with initial velocity $(\nu_0,\psi_0,\alpha_0)=(0.3,\pi/2,-\pi/3)$.
In all plots, the radiation source is the green line with radius $x_\star=1.5$, while the gray surface is the critical hypersurface, and the black sphere is the BH. The arrow represents the polar axis.}  
\label{fig:Fig11}
\end{figure*}

In Fig. \ref{fig:Fig11}, we report different examples of test particle trajectories for various quadrupole moment values $q=-4,-1,1,4$. The test particle can conclude its motion either at infinity or on the critical hypersurface/radiation emitting source \cite{Bini2009,Bini2011,Bini2015,DeFalco20183D,Bakala2019}. In addition, the test particle when reaches the critical hypersurface can be dragged towards the equatorial plane showing the so-called latitudinal drift, due to the interplay between tidal gravitational forces and PR effect. There is also another interesting configuration, where the test particle moves on suspended orbits, without being drifted down to the equatorial plane (see blue line in the bottom right panel of Fig. \ref{fig:Fig11}). These plots graphically show the critical hypersurface shape, which we have already analysed in detail in Sec. \ref{sec:PCH}. We note that for lower luminosities and angular velocities, the critical hypersurface size is very close to the emitting surface; while for higher values of these two quantities the dimension of the critical region increases. We have also displayed that if the test particle starts its motion in the equatorial plane, it will stay over there forever without moving out (see orange line in the bottom right panel of Fig. \ref{fig:Fig11}).  

\subsection{Multiplicity of critical hypersurface}
\label{sec:MCH}
Equation (\ref{eq:CH}) may exhibit three different critical hypersurface solutions (see Refs. \cite{Bini2011,Bini2015,Bakala2019}, for further details). In order to easily check the multiplicity of the solutions, we approximate Eq. (\ref{eq:CH}) through the following function (see Ref. \cite{Bakala2019}, for details)
\begin{equation}
\frac{a_3x^3+a_2x^2+a_1x+a_0}{x^5}.
\end{equation}
The polynomial of third order reads explicitly as
\begin{eqnarray}
a_3&&=1-A,\\
a_2&&=-\frac{b^2}{1-y^2}-1,\\
a_1&&=\frac{A(2y^2+3b^2-2)}{2\left(1-y^2\right)} -\frac{q(3y^4-4y^2+1)}{5\left(1-y^2\right)}\nonumber\\
&&\quad-\frac{3(y^2-4b^2-1)}{2\left(1-y^2\right)},\\
a_0&&=-\frac{3b^2}{1-y^2}\left(2A+\frac{13}{2}\right)-\frac{4q}{5}\left(y^2-\frac{1}{3}\right)-\frac{3}{2}.
\end{eqnarray}
The multiplicity of solutions of a polynomial equation of third order is achieved through the discriminant criterion $\Delta_{\rm III}$, defined as follows \cite{Bakala2019}
\begin{equation}
\Delta_{\rm III}=18a_3a_2a_1a_0-4a_2^3a_0+a_2^2a_1^2-4a_3a_1^3-27a_3^2a_0^2.
\end{equation}
If $\Delta_{\rm III}<0$ we have only one real solution, if $\Delta_{\rm III}=0$ we have two different real solutions with one counted twice, and finally if $\Delta_{\rm III}>0$ we have three distinct real solutions. Only one of this solution is physical, because we have one inside the emitting surface (unphysical), one close to the emitting surface (physical), and the last one very far from the emitting surface (unphysical) \cite{Bini2011,Bakala2019}.

\subsection{Suspended orbits}
\label{sec:SO}
The test particle could move on circular orbits bounded on the critical hypersurface at constant height $y\neq0$ (off-equatorial plane), without the action of the latitudinal drift mechanism (see Refs. \cite{DeFalco20183D,Bakala2019}, for further details). To obtain such configurations, the test particle must touch the critical hypersurface with the following conditions: $\alpha=0,\pi$, $\nu=\cos\beta$, and $d\psi/d\tau=0$. Vanishing Eq. (\ref{EoM2}), it is possible to determine the value of $\psi$, by solving this implicit equation \cite{Bakala2019}:
\begin{equation}\label{eq:SO}
\begin{aligned}
&\gamma\sin\psi\left[a(\boldsymbol{n})^{\hat y}+\nu^2k(\varphi,\boldsymbol{n})^{\hat y}\right]\\
&+\frac{A\sqrt{1-y^2}e^{-\lambda}(1-\nu^2\sin\psi)\nu\cos\psi}{M \sqrt{(x^2-y^2)[M^2(x^2-1)(1-y^2)-b^2e^{4\mu}}}=0.
\end{aligned}
\end{equation}
In the special case $b=0$, the condition for suspended orbits is achieved for $\psi=\pm\pi/2$ \cite{DeFalco20183D,Bakala2019}. The value of $\psi$ strongly depends on emitting surface location $x_\star$, angular velocity $\Omega_\star$, and compact object quadrupole moment $q$. 
\begin{figure}[htbp]  
\begin{center}
\includegraphics[scale=0.38]{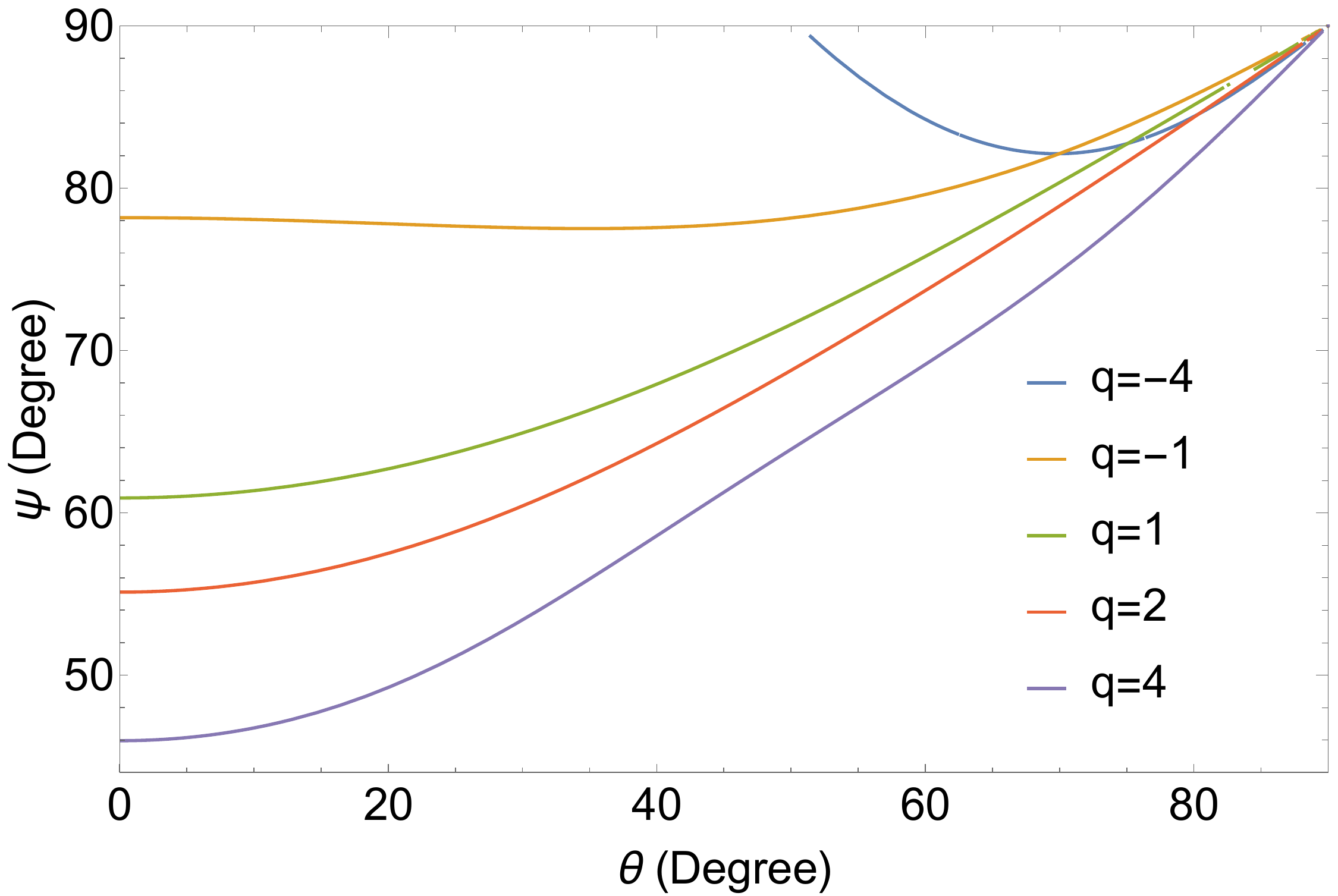}
\end{center}
\caption{Angle $\psi$ in terms of the polar angle $\theta$ for $q=-4,-1,1,2,4$, set $A=0.9$, $\Omega_\star=0.017$, and $x_\star=1.5$.}  
\label{fig:Fig10}
\end{figure}

In Fig. \ref{fig:Fig10} we show the angle $\psi$ at which the test particle should touch the critical hypersurface to reach the fixed height $y$, and moving in such plane on a circular orbit. We note that there is an interesting effect depending on the sign of $q$. For $q\ge-1$ it is possible to have suspended orbits at every height, while for $q<-1$, this is possibile only for a small range close to the equatorial plane. In the former case there is a perfect balance among all forces at any fixed height; while in the last case the action of the strong tidal forces (generated by the quadrupole moment) do not permit to have stable circular orbits towards the poles. The equatorial suspended orbits always exist, because for the latitudinal drift effect the test particle is forced to end its motion over there. 

\subsection{Stability of critical hypersurfaces}
\label{sec:SCH}
The general relativistic PR effect is expected to be relevant for high luminosity compact objects as it contributes determining the matter motion in their close vicinity; the inner regions of accretion disks and coronas \cite{Inogamov1999,Fabian2015}, or the expanding NS photosphere during bright X-rays bursts are just a few examples \cite{Lewin1993}. Our present treatment entails approximations (such 
as spherical symmetry of the emitter, or the absence of frame dragging) that  limit its applicability to astrophysical problems. Nevertheless stability of the critical hypersurfaces is an important feature to assess. De Falco and collaborators \cite{DeFalco2019ST} have introduced a new, simple, and elegant method employing the Lyapunov functions\footnote{A Lyapunov function, associated to a dynamical system $\boldsymbol{\dot{x}}=\boldsymbol{f}(\boldsymbol{x})$, is a real valued smooth map $\Lambda: \boldsymbol{\chi}\equiv(\nu,\alpha,\psi,x,y)\in\mathcal{D}\equiv[0,1]\times[0,2\pi]\times[0,\pi]\times[1,\infty]\times[-1,1]\to\Lambda(\boldsymbol{\chi})\in\mathbb{R}$ for a point $\boldsymbol{\chi_0}$, such that $\boldsymbol{f}(\boldsymbol{\chi_0})=\boldsymbol{0}$, if it fulfills the conditions \cite{DeFalco2019ST}
\begin{eqnarray}  
{\rm (I)}&&\quad \Lambda(\boldsymbol{\chi})>0,\quad \forall \boldsymbol{\chi}\in \mathcal{D}\setminus\left\{\boldsymbol{\chi_0}\right\},\label{eq:lia1}\\
{\rm (II)}&&\quad \Lambda(\boldsymbol{\chi_0})=0,\label{eq:lia2}\\ 
{\rm (III)}&&\quad \dot{\Lambda}(\boldsymbol{\chi})\equiv\nabla\Lambda(\boldsymbol{\chi})\cdot \boldsymbol{f}(\boldsymbol{\chi})\le0 ,\quad \forall \boldsymbol{\chi}\in \mathcal{D}.\label{eq:lia3}
\end{eqnarray}
This implies that the point $\boldsymbol{\chi_0}$ is stable.}, which have deep mathematical and physical meanings. Three different Lyapunov functions have been proposed, which are: kinetic energy $\mathbb{K}$, angular momentum $\mathbb{L}$, and Rayleigh potential $\mathbb{F}$ \cite{DeFalco2019ST}. They are all defined relative to the critical hypersurface.

\begin{figure*}
\begin{center}
\vbox{
\hbox{
\includegraphics[trim=1cm 2cm 0cm 1cm,scale=0.34]{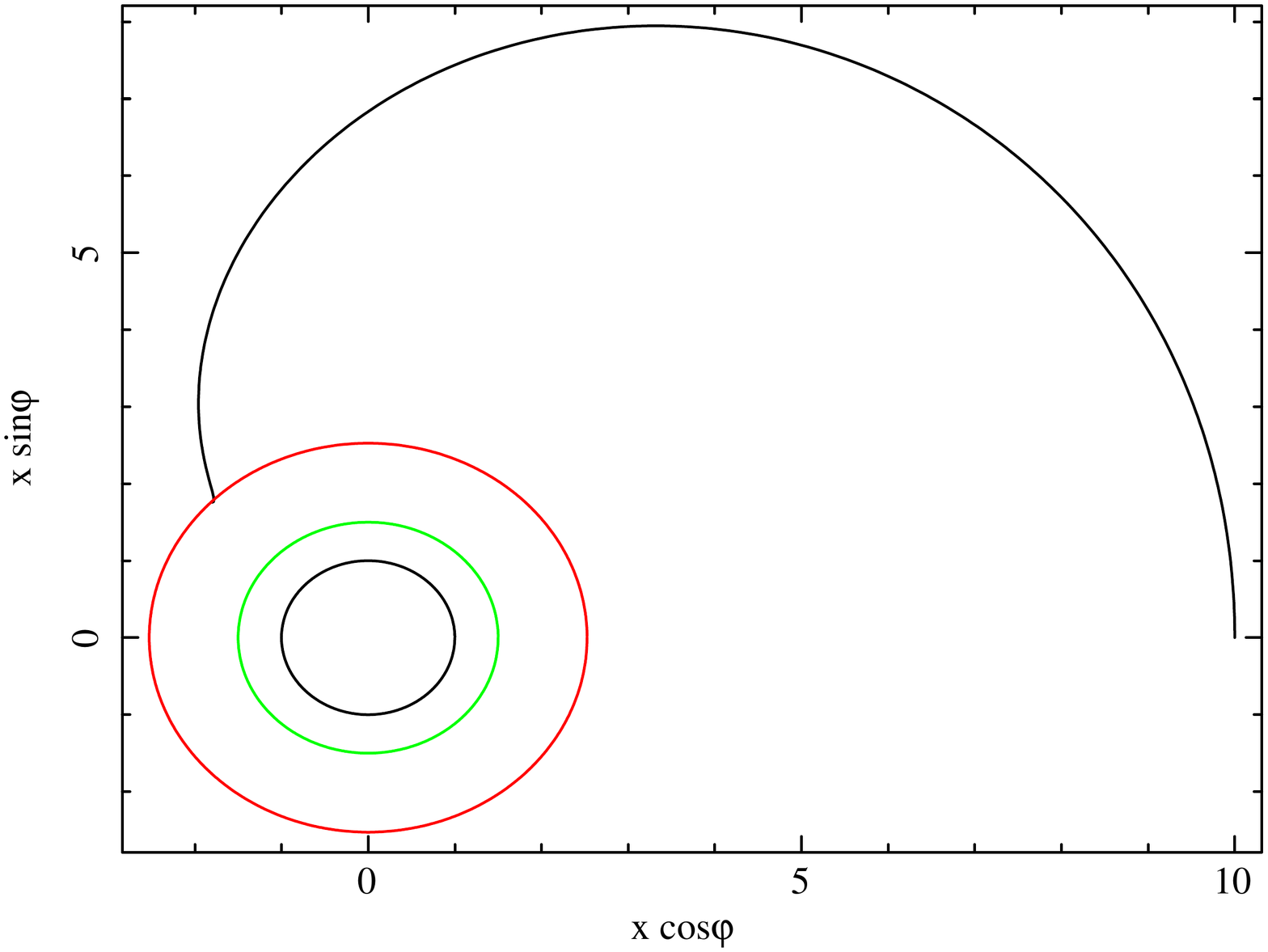}
\hspace{0.1cm}
\includegraphics[trim=1cm 2cm 0cm 1cm,scale=0.34]{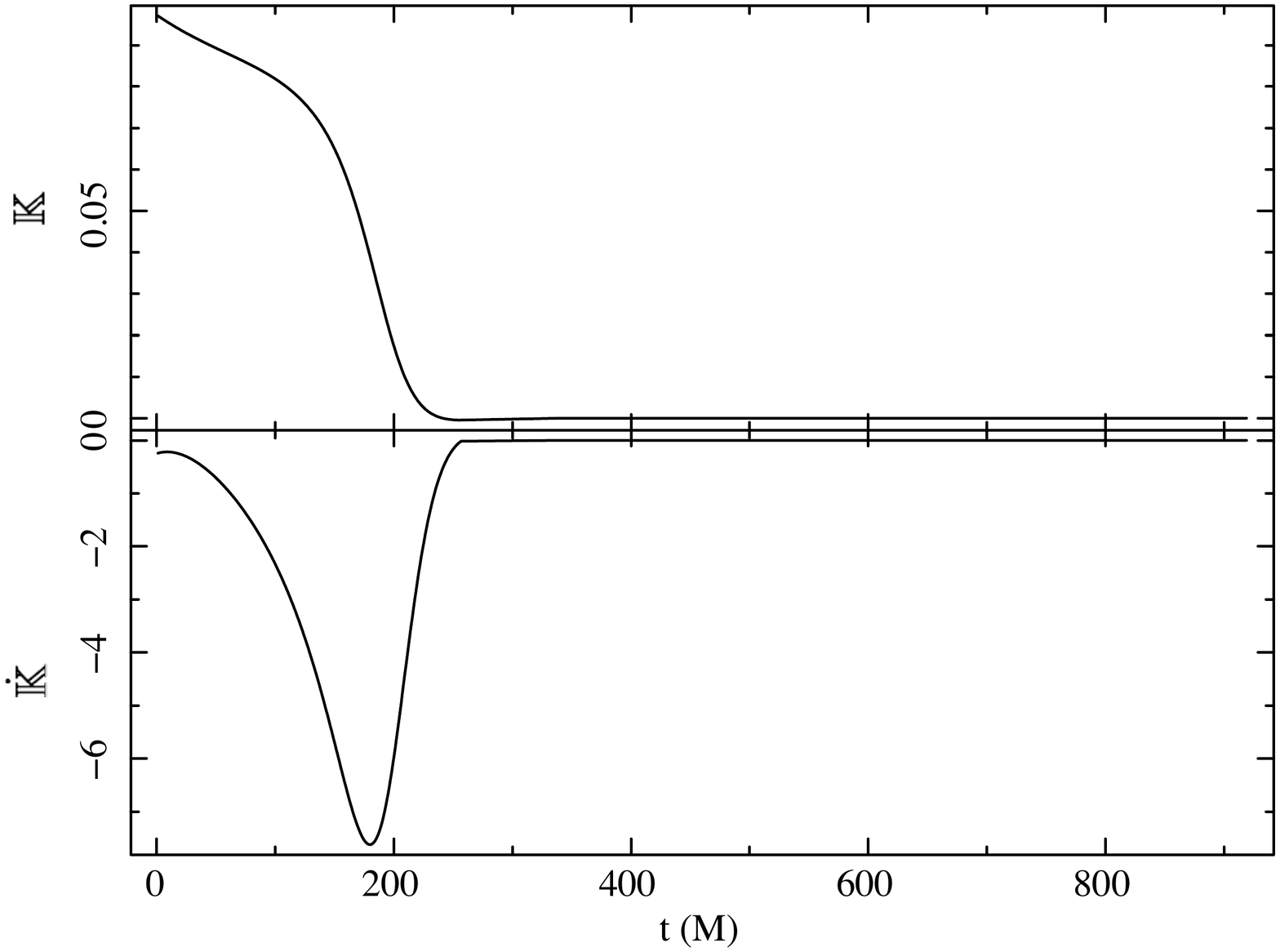}}
\vspace{0.5cm}
\hbox{
\includegraphics[trim=1cm 2cm 0cm 1cm,scale=0.34]{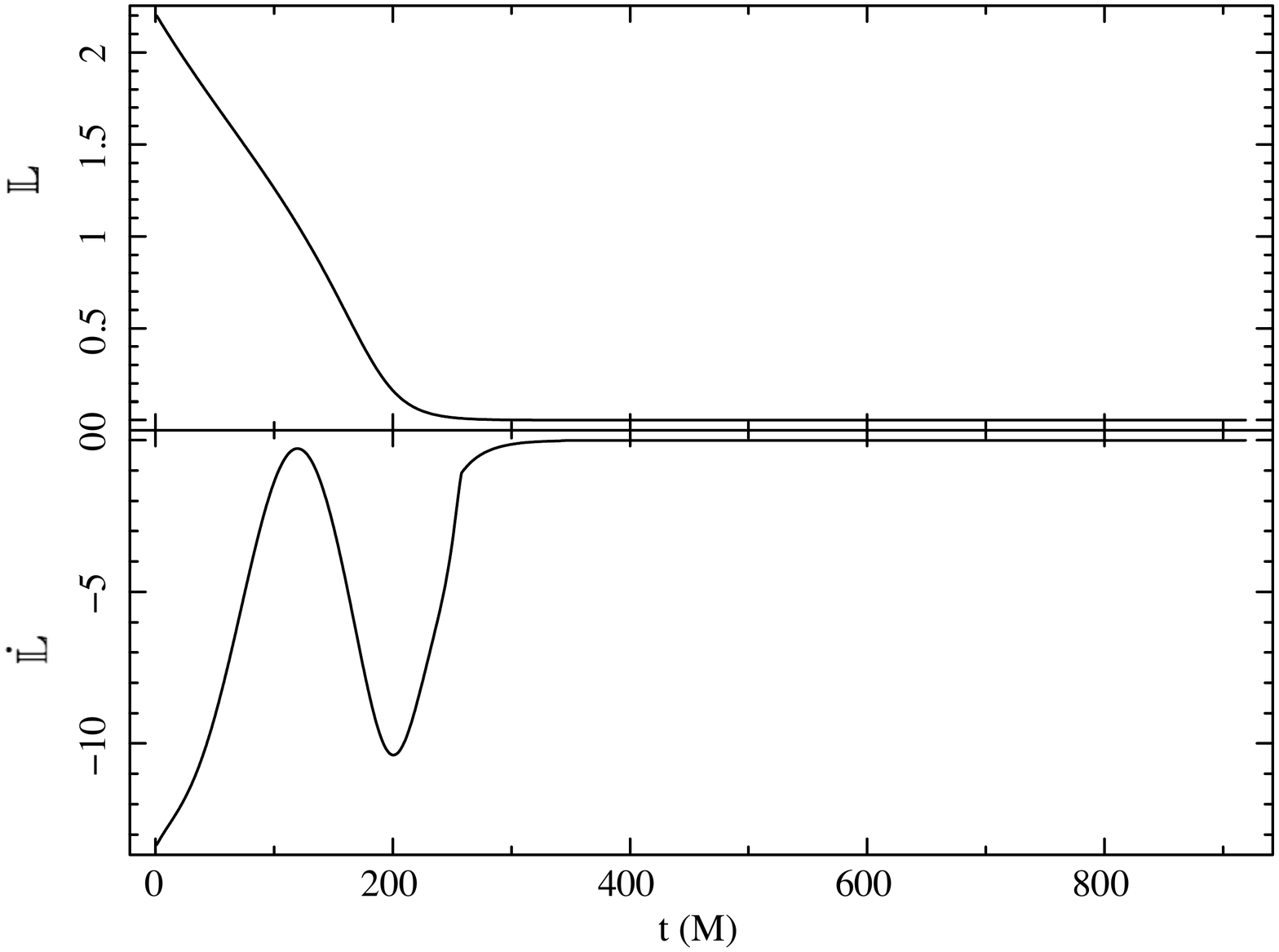}
\hspace{0.1cm}
\includegraphics[trim=1cm 2cm 0cm 1cm,scale=0.34]{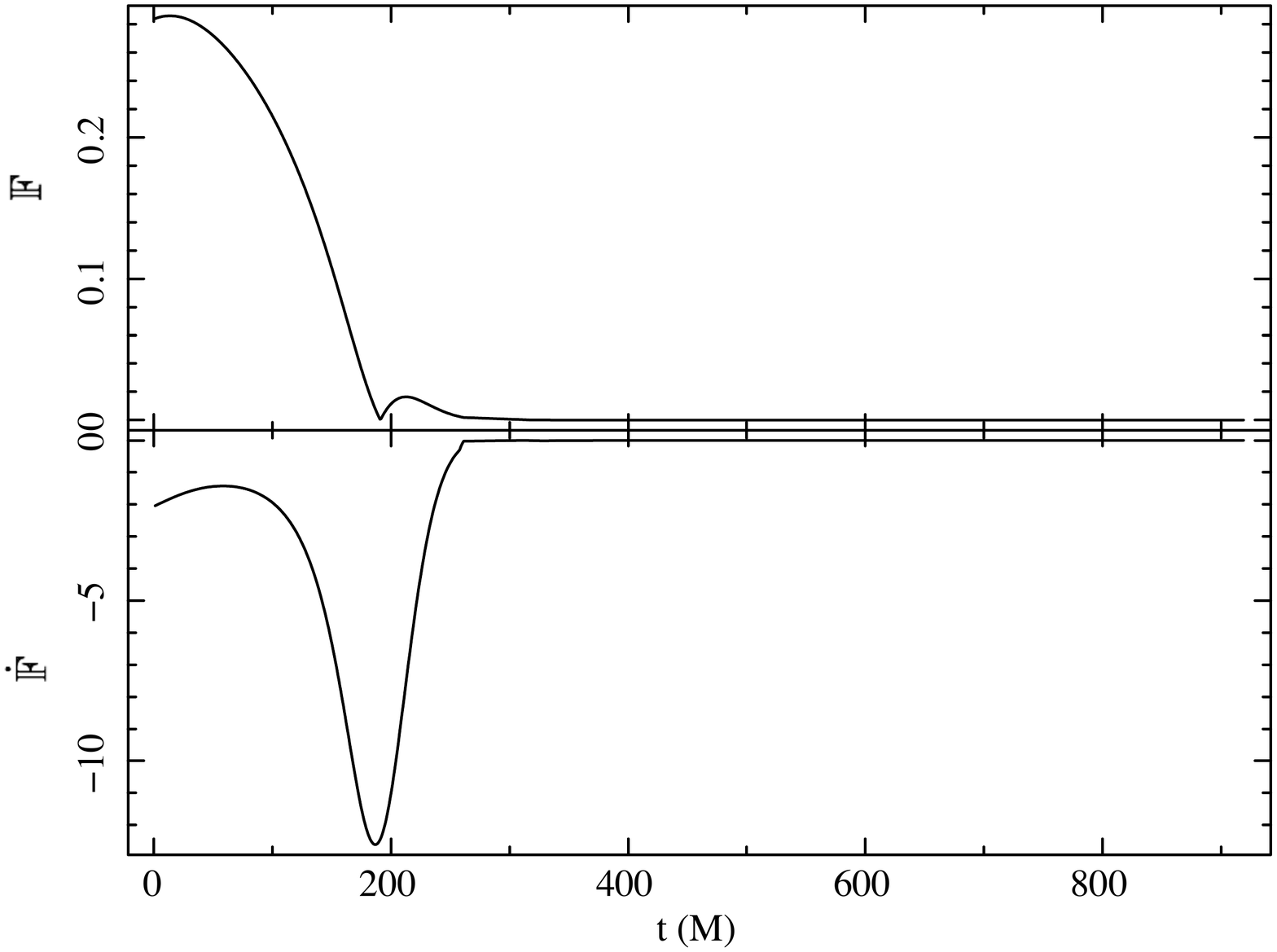}}}
\end{center}
\caption{Test particle orbit and the related three Lyapunov functions, which graphically prove the stability of the critical hypersurface. Upper left panel: test particle moving around a BH in the equatorial plane (i.e., $\psi_0=\pi/2$ and $y=0$) with quadrupole moment $q=3$, luminosity parameter $A=0.6$, and photon impact parameter $b=0$. The test particle starts its motion at the position $(r_0,\varphi_0)=(10,0)$ with velocity $(\nu_0,\alpha_0)=(0.2,0)$. The black, green, and red circles represent BH event horizon $x_H=1$, emitting surface $x_\star=1.5$, and critical hypersurface $x_{\rm crit}=2.53$, respectively. The three Lyapunov functions (\ref{eq:l1}) --  (\ref{eq:l3}) are plotted in arbitrary units together with their $\tau$-derivatives in terms of the coordinate time $t$.}  
\label{fig:Fig12}
\end{figure*}

Defined $\nu_{\rm crit}$ as the velocity of the test particle on the critical hypersurface, the Lyapunov functions are\footnote{We set to unity some constants, and from the function $\mathbb{F}$ we did not considered the term $\tilde{\sigma}\Phi^2$, because its absence does not alter any of the proprieties of the Lyapunov function.} \cite{DeFalco2019ST}
\begin{eqnarray}
\mathbb{K}&=&\frac{\left|\nu^2-\nu^2_{\rm crit}\right|}{2}+\left[\frac{A}{M}-1\right]\left[\frac{1}{x+1}-\frac{1}{x_{\rm crit}+1}\right],\label{eq:l1}\\
\mathbb{L}&=&\left[(x+1)\nu\sin\psi\cos\alpha-(x_{\rm crit}+1)\nu_{\rm crit}\right],\label{eq:l2}\\
\mathbb{F}&=&\left|\lg\left(\frac{E_{\rm crit}}{E}\right)-\lg\left(\frac{E(\boldsymbol{U})}{E}\right)\right|,\label{eq:l3}
\end{eqnarray}
where $E_{\rm crit}$ is the energy $E(\boldsymbol{U})$ evaluated on the critical hypersurface. The formal proof of the stability can be reconstructed by following the same strategy of Ref. \cite{DeFalco2019ST}, and employing the results contained in Appendix \ref{sec:WFL}. Here, we do not report the explicit calculations, but we limit only to graphically show it, see Fig. \ref{fig:Fig12}. 

\section{Conclusions}
\label{sec:end}
We have developed the full general relativistic 3D treatment of the motion of a test particle around a static and non-spherical compact object, described by the Erez-Rosen metric, and affected by the radiation field from a spherical and rigidly rotating emitting surface located outised the compact object, generating radiation pressure and PR drag force (see Sec. \ref{sec:dynamics}). The Erez-Rosen spacetime belongs to the Weyl class, containing all the axisymmetric and static vacuum solutions of the Einstein's field equations. A non-spherical source can be described through a mass multipole moment expansion of the metric, but we limit our consideration to the quadrupole moment. This represents the threshold under which all the metrics of Weyl class are equivalent through an opportune change of variables (see Sec. \ref{sec:gen_met}). In particular we decided to employ the Erez-Rosen metric because it has been already used in the 2D description of the PR effect in the equatorial plane \cite{Bini2015}, in order to compare the results of the 3D case (see Sec. \ref{sec:ER_met}). 

This work offers also a great opportunity to investigate the pure geometrical aspects of the Eerez-Rosen metric in the 3D space, never treated so far in the literature, although there are plenty of works in the 2D case. The role played by the kinematical quantities is fundamental for the geometrical description of this spacetime (see Table \ref{tab:SOq}). Through the weak field approximation (see Appendix \ref{sec:WFL}), we have figured out that the radial (polar) components describe the radial (polar) curvature in the three directions $x,y,\varphi$. From their Taylor expansions in $x$, it is possible to see how these terms behave in the Schwarzschild limit $q=0$. In particular, we have learned that at the fourth-order in $x$, $a(\boldsymbol{n})^{\hat x}\equiv k(x,\boldsymbol{n})^{\hat x}$, and $a(\boldsymbol{n})^{\hat y}\equiv k(x,\boldsymbol{n})^{\hat y}$; while at the third order in $x$, $k(y,\boldsymbol{n})^{\hat x}\equiv k(\varphi,\boldsymbol{n})^{\hat x}$, and $k(y,\boldsymbol{n})^{\hat y}\equiv k(\varphi,\boldsymbol{n})^{\hat y}$; and the quadrupole moment occurs as a fourth- and fifth-order effect in the geometrical and radiation part, respectively. 

The dynamical equations have a very flexible structure to the changes of geometrical backgrounds and radiation field's descriptions. Indeed, we have combined the Erez-Rosen metric with the actual treatment of the radiation field framed in the Kerr metric  \cite{DeFalco20183D,Bakala2019}. Besides to an obvious change in the functional form of the kinematical quantities, the radiation field differs to the previous description only for the parameter $\Phi^2$, which encodes the geometrical aspects of the background metric.

This dynamical system exhibits the presence of a critical hypersurface, which changes its form in terms of luminosity parameter $A$, emitting surface radius $x_\star$ and angular velocity $\Omega_\star$, and the quadrupole moment $q$, see Sec. \ref{sec:CH}, and Figs. \ref{fig:Fig2} -- \ref{fig:Fig8}, for further details.

Selected test particle orbits have been displayed (see Fig. \ref{fig:Fig11}), which show the characteristic latitudinal drift of the test particle towards the equatorial plane (see Sec. \ref{sec:TPO}). This is caused by the interplay among tidal forces and PR drag force, which find their equilibrium in the equatorial plane. The test particle has two destinies: either escaping at infinity or ending its motion on the critical hypersurface. The test particle can also end its motion on suspended orbits on the critical hypersurface (see Sec. \ref{sec:SO}). We have calculated the condition for achieving such configurations and we have noted: for $q<-1$ it is possible to have suspended orbits only in a close neighborhood of the equator, due to the presence of strong tidal forces towards the poles; while for $q\ge-1$ it is possible to have such configurations at all heights, because there is a perfect balance among all forces.  

The implicit equation (\ref{eq:CH}) describing the critical hypersurface can admit from one to three solutions, depending on the set of initial parameters. We have found an accurate approximation of such equation in terms of a polynomial of third order. Calculating its discriminant, it would be possible to easily infer the multiplicity of solutions (see Sec. \ref{sec:MCH}). However, the solutions are composed by two unphysical (one located inside the emitting surface and another one very far from the compact object), and only one physically acceptable.  

Finally in Fig. \ref{fig:Fig12}, we have graphically proven that the critical hypersurface is a basin of attraction and the equatorial ring is a stable attractor, employing a new approach based on the Lyapunov theory (see Sec. \ref{sec:SCH}). 

The non-spherical shape of the compact objects permits to have critical hypersurfaces very close to the event horizon, which might be advantageously exploited to develop new tests of the Einstein's theory in strong field regimes. Relevant quadrupole deformations are also caused by the rotation of the compact object, as it also occurs in the Kerr case. Hence, in the next work we would like to include the rotation and quadrupole moment in a Hartle-Thorne metric-like \cite{Hartle1967,Hartle1968}, to see how this new general relativistic framework couples with the radiation field.

\section*{Acknowledgements}
V.D.F. thanks the Silesian University in Opava and Gruppo Nazionale di Fisica Matematica of Istituto
Nazionale di Alta Matematica for the support. V.D.F. is grateful to Professor Hernando Quevedo and Doctor Daniela Pugliese for the useful discussions on the Erez-Rosen metric. V.D.F. is grateful to Professor Luigi Stella for the useful discussions. P.B. acknowledges the Czech Science Foundation (GA{\"A}R) grant GA{\"A}R 17-16287S and internal grant of Silesian University in Opava SGS/13/2019.

\appendix

\section{Weak field Limit}
\label{sec:WFL}
It is interesting to consider the weak field limit of Eqs. (\ref{EoM1}) -- (\ref{EoM6}), i.e., considering $x\to\infty$ keeping all terms linear in $q$. Therefore, we obtain
\begin{eqnarray} \label{eq:WFKQ}
&&a(\boldsymbol{n})^{\hat x}\to
\frac{1}{x^2}\left(1-\frac{1}{x}+\frac{6 q y^2-2 q+15}{10x^2}\right)+O\left(\frac{1}{x^5}\right),\nonumber\\
&&k(x,\boldsymbol{n})^{\hat x}\to\frac{1}{x^2}\left(1-\frac{1}{x}+\frac{6 q y^2-2 q+15}{10x^2}\right)+O\left(\frac{1}{x^5}\right),\nonumber\\
&&k(y,\boldsymbol{n})^{\hat x}\to-\frac{1}{x}\left[1+ \frac{2}{x}-\frac{5}{2x^2}\right.\nonumber\\
&&\hspace{1.8cm}\left.+\frac{0.8(q y^2-1/3q+3.75)}{x^3}\right] +O\left(\frac{1}{x^5}\right),\nonumber\\
&&k(\varphi,\boldsymbol{n})^{\hat x}\to-\frac{1}{x}\left[1+\frac{2}{x}-\frac{5}{2x^2}\right.\nonumber\\
&&\hspace{1.8cm}\left.+\frac{1}{x^3}\left(\frac{4 q y^2}{5}-\frac{4 q}{15}+3\right)\right]+O\left(\frac{1}{x^5}\right),\nonumber\\
&&a(\boldsymbol{n})^{\hat y}\to-\frac{2qy}{5x^4}\sqrt{1-y^2}+O\left(\frac{1}{x^5}\right),\nonumber\\
&&k(x,\boldsymbol{n})^{\hat y}\to-\frac{2qy}{5x^4} \sqrt{1-y^2}+O\left(\frac{1}{x^5}\right),\\
&&k(y,\boldsymbol{n})^{\hat y}\to\frac{y}{x \sqrt{1-y^2}}\left(1+\frac{1}{x}+\frac{1}{x^3}\right.\nonumber\\
&&\hspace{1.8cm}\left.+\frac{7 q -9 q y^2-15}{15x^3}\right)+O\left(\frac{1}{x^5}\right),\nonumber\\
&&k(\varphi,\boldsymbol{n})^{\hat y}\to -\frac{y}{x \sqrt{1-y^2}}\left(1+\frac{1}{x}+\frac{1}{x^3}\right.\nonumber\\
&&\hspace{1.8cm}\left.+\frac{5 q-3 q y^2+15}{15x^3}\right)+O\left(\frac{1}{x^5}\right)\nonumber.
\end{eqnarray}
We note that \emph{the quadrupole moment $q$ is an effect of fourth-order in $x$}. In addition, such limit gives important information on the Erez-Rosen spacetime, and the involved kinematical quantities. Indeed, we find
\begin{equation}
a(\boldsymbol{n})^{\hat x}\equiv k(x,\boldsymbol{n})^{\hat x},\qquad a(\boldsymbol{n})^{\hat y}\equiv k(x,\boldsymbol{n})^{\hat y},
\end{equation}
valid only up to the fourth-order in $x$, because from the fifth-order on, some differences start to appear in the two Taylor-expansions. We obtain another important result, 
\begin{equation}
k(y,\boldsymbol{n})^{\hat x}\equiv k(\varphi,\boldsymbol{n})^{\hat x},\qquad k(y,\boldsymbol{n})^{\hat y}\equiv k(\varphi,\boldsymbol{n})^{\hat y},
\end{equation}
valid only up to the third-order in $x$, because at higher orders there are discrepancies. These limits show an important feature of the Erez-Rosen metric, that exhibits an high degree of symmetry in its geometrical structure.

Now, let us consider the weak filed limit of the test particle velocity field, given by Eqs. (\ref{eq:velocitycomp}), (keeping always linear terms in $q$), i.e.,
\begin{eqnarray}\label{eq:WFV}
\gamma\nu\sin\psi\sin\alpha&&=\dot{x}\left\{1+\frac{1}{x}+\frac{1}{2x^2}\right.\nonumber\\
&&\left.+\frac{1}{30x^3}\left(6q y^2-2q+15\right)+O\left(\frac{1}{x^4}\right)\right\},\nonumber\\
\gamma\nu\cos\psi&&=\frac{\dot{y}x}{\sqrt{1-y^2}}\left\{1+\frac{1}{x}\right.\nonumber\\
&&\left.+\frac{q(3 y^2-1)}{15x^3}+O\left(\frac{1}{x^4}\right)\right\},\\
\gamma\nu\cos\alpha\sin\psi&&=\dot{\varphi}x\sqrt{1-y^2}\left\{1+\frac{1}{x}\right.\nonumber\\
&&\left.+\frac{q(3y^2-1)}{15 x^3}+O\left(\frac{1}{x^4}\right)\right\},\nonumber
\end{eqnarray}
where the dot stands for the derivative with respect to the proper time $\tau$. We consider also the weak field limit of $U^t$, in order to transform all the derived quantities with respect to the coordinate time $t$. Therefore, Eq. (\ref{eq:velocitycomp}) becomes
\begin{equation}\label{eq:WFTIME}
\begin{aligned}
U^t\equiv \dot{t}&=\gamma\left[1-\frac{1}{x}+\frac{1}{2x^2}\right.\\
&\left.-\frac{1}{30x^3}\left(6qy^2-2q+15\right)\right]+O\left(\frac{1}{x^4}\right).
\end{aligned}
\end{equation}

Combining such limits Eqs. (\ref{eq:WFKQ}), (\ref{eq:WFV}), and (\ref{eq:WFTIME}), we can obtain the weak field approximation of the test particle acceleration components. It the slow motion approximation (keeping only linear terms in $\nu$) such components assume the following form
\begin{eqnarray} \label{eq:WFA}
a(\boldsymbol{U})^{\hat x}&&=\frac{d}{dt}\left\{\frac{dx}{dt}\dot{t}\left[1+\frac{1}{x}+\frac{1}{2x^2}+\frac{1}{30x^3}\left(6q y^2-2q+15\right)\right]\right\}\dot{t}\nonumber\\
&&+\frac{1}{x^2}\left(1-\frac{1}{x}+\frac{6 q y^2-2 q+15}{10x^2}\right)+O\left(\frac{1}{x^5}\right),\nonumber\\
a(\boldsymbol{U})^{\hat y}&&=\frac{d}{dt}\left\{\frac{dy}{dt}\frac{\sqrt{1-y^2}}{x}\dot{t}\left[1+\frac{1}{x}+\frac{q(3 y^2-1)}{15x^3}\right]\right\}\dot{t}\nonumber\\
&&-\frac{2qy}{5x^4}\sqrt{1-y^2}+O\left(\frac{1}{x^5}\right),\\
a(\boldsymbol{U})^{\hat \varphi}&&=\frac{d}{dt}\left\{\frac{d\varphi}{dt}\dot{t}\frac{1}{x\sqrt{1-y^2}}\left[1+\frac{1}{x}+\frac{q(3y^2-1)}{15 x^3}\right]\right\}\dot{t}.\nonumber
\end{eqnarray}

Instead for the radiation force components, we calculate the weak field limit of $\tilde{\sigma}[\Phi E(\boldsymbol{U})]^2$, which is
\begin{equation} 
\tilde{\sigma}[\Phi E(\boldsymbol{U})]^2=A f(x,y,q)\gamma^2 [1-\nu\sin\psi\cos(\alpha-\beta)]^2,
\end{equation}
where
\begin{equation}
\begin{aligned}
f(x,y,q)&=\frac{1}{M^2x^2}
+\frac{\left(b^2-2 M^2 y^2+2 M^2\right)}{2x^4 M^4 \left(1-y^2\right)}
-\frac{2 b^2}{M^4 x^5\left(1-y^2\right)}\\
&-\frac{-15 b^4+200 b^2 M^2 y^2-200 b^2 M^2+20 M^4 q y^8}{40M^6x^6 \left(1-y^2\right)^2}\\
&+\frac{-64 M^4 q y^6+72 M^4 q y^4-32 M^4 q y^2+4 M^4 q}{40M^6x^6 \left(1-y^2\right)^2}\\
&+\frac{-40 M^4 y^4+80 M^4 y^2-40 M^4}{40M^6x^6 \left(1-y^2\right)^2}+O\left(\frac{1}{x^7}\right).
\end{aligned}
\end{equation}
In the radiation force \emph{the quadrupole moment $q$ is an effect of fifth-order in $x$}, being therefore weaker than the geometrical terms. In the slow motion limit the radiation field components read as
\begin{eqnarray}
F_{\rm (rad)}^{\hat x}&&=A f(x,y,q)\left\{\sin\beta-\nu\sin\psi[\cos(\alpha-\beta)\sin\beta\right.\nonumber\\
&&\left.+\sin\alpha]\right\},\\
F_{\rm (rad)}^{\hat y}&&=-A f(x,y,q)\nu\cos\psi,\\
F_{\rm (rad)}^{\hat \varphi}&&=A f(x,y,q)\left\{\cos\beta-\nu\sin\psi[\cos(\alpha-\beta)\cos\beta\right.\nonumber\\
&&\left.+\cos\alpha]\right\},
\end{eqnarray}
where
\begin{eqnarray}
\cos\beta&&=\frac{b}{x\sqrt{1-y^2}}\left[1-\frac{2}{x}+\frac{5}{2x^2}\right.\nonumber\\
&&\left.-\frac{6 q y^2-2 q+45}{15x^3}
\right]+O\left(\frac{1}{x^5}\right)\\
\sin\beta&&=1-\frac{b^2}{x^2(1-y^2)}\left[\frac{1}{2}-\frac{2}{x}+\frac{1}{2x^2}\left(9+\frac{b^2}{4(1-y^2)}\right)\right.\nonumber\\
&&\left.-\frac{\left(15 b^2-6 q y^4+8  q y^2-2  q-120  y^2+120 \right)}{15x^3(1-y^2)}\right]\nonumber\\
&&+O\left(\frac{1}{x^6}\right).
\end{eqnarray}
\bibliography{references}

\end{document}